\documentclass[a4paperal]{i3m}

\usepackage{graphicx}
\usepackage{siunitx}
\usepackage{amsfonts}
\usepackage{amsmath}
\usepackage{wasysym}
\usepackage{algorithm}
\usepackage{algpseudocode}
\usepackage{ragged2e}

\conference{emss}

\title{Tournament versus Circulant: On Simulating 7-Species Evolutionary Spatial Cyclic Games with Ablated Predator-Prey Networks as Models of 
Biodiversity}
\author{Dave Cliff}

\affil{Intelligent Systems Laboratories, School of Engineering Mathematics and Technology, University of Bristol, U.K.}

\authnote{Email address: csdtc@bristol.ac.uk}

\runningauthor{Dave Cliff}

\confyear{2023}

\begin{document}

\begin{frontmatter}
\maketitle
%
\begin{abstract}
\justifying
Computer simulations of minimal population-dynamics models have long been used to explore questions in ecosystems coexistence and species biodiversity, via simple agent-based models of three interacting species, referred to as  $R$, $P$, and $S$, 
and where individual agents compete with one another in predator/prey contests that are determined by the cyclic dominance rules of the Rock-Paper-Scissors game. Recent publications have explored the dynamics of five-species models, based on the Rock-Paper-Scissors-Lizard-Spock (RPSLS) game. A 2022 paper by 
Zhong  et al.\ reported simulation studies of species coexistence in spatial RPSLS systems in which one or more directed edges are ablated from the five-vertex tournament digraph defining the RPSLS game: 
Zhong  et al.\ showed that the ablation of a single inter-species interaction can lead to a collapse in biodiversity. In this paper I present first results from simulation studies of evolutionary spatial cyclic games where there are seven species, but where each species is still in its own local five-species RPSLS-like interaction network: the dominance networks I use for this are a subset of the 
$n$-node $k$-regular circulant digraphs $D(n,\Omega)$ for odd-numbered $n$ and $|\Omega|=2$.
My results indicate that 
Zhong  et al.’s results are due to the specific fully-connected tournament dominance network used in their RPSLS model: when other, equally realistic, RPSLS-style dominance networks are used instead, no such sudden collapse in biodiversity occurs. The Python source-code used for the work reported here is freely available on GitHub.
\end{abstract}
\end{frontmatter}

\section{Introduction}
\label{sec:intro}

In theoretical biology research there is a long tradition of using computer simulations of minimal population-dynamics models to explore questions in ecosystems coexistence and species biodiversity via simple agent-based models of three interacting species, referred to as $R$, $P$, and $S$, and where individual agents compete with one another in life-or-death predator/prey contests that are determined by the cyclic dominance rules of the Rock-Paper-Scissors (RPS) game. 
Much of this research has been focused on evolutionary spatial cyclic games (ESCGs), where each agent occupies a specific location on a lattice or grid of cells, competes only with its immediate neighbours, and is able to move around the lattice over time. In these minimal RPS models of ecosystems, each `species' forever plays one of the three available moves, and the questions of interest concern how many species can co-exist, and what is the asymptotic steady-state frequency of each species, under differing circumstances. 
More recent publications have explored the dynamics of slightly richer five-species ECSG systems, based on an extension of RPS known as Rock-Paper-Scissors-Lizard-Spock (RPSLS). A recent paper by  \cite{zhong_etal_2022_ablatedRPSLS} reported simulation studies of species coexistence in spatial RPSLS systems  
in which one or more of the species interactions are ablated from the game's dominance network, i.e. one or more directed edges are deleted from the five-vertex directed graph (or {\em digraph}) defining the dominance relationships for the RPSLS game (Zhong et al.\ refer to the dominance network as the ``interaction structure''). Zhong  et al. showed that the ablation of a single inter-species interaction can lead to a collapse in biodiversity, with the number of coexisting species falling by more than 50\%, with the final outcome sometimes being only two or fewer species remaining.  However, because the complete RPSLS digraph has a total of ten directed edges, Zhong  et al.\/'s deletion of a single edge amounts to ablating 10\% of the entire network of inter-species interactions, which would amount to a fairly major change in any ecosystem and is such a large perturbation that   it is arguably no surprise that extinctions then follow. 

In this paper I present first results from simulation studies of  RPSLS-style ECSGs where there are seven species, but where each species is still in its own local five-species RPSLS interaction network. This is novel, and involves recognising that the dominance networks for the 3-species RPS system and the 5-species RPSLS system are both, in the language of graph theory (explained further in Section~\ref{sec:graphtheory}), {\em k-regular circulant} digraphs. Other authors have referred to these dominance networks as {\em tournament} digraphs, which is strictly correct, but these tournament digraphs are a subset . A {\em tournament} digraphs is one where every vertex $v_i$ (representing species $S_i$) is joined by a single directed edge to every other vertex $v_{j\neq i}$ in the network -- that is, if we let $N_S$ denote the number of species in the model ecosystem, in the $N_S$$=$$3$ (RPS) and $N_S$$=$$5$ (RPSLS) models, {\em every} species is in a direct predator/prey relationship with {\em all}\/ other species: for any pair of species $S_i$ and $S_{j\neq i}$, either $S_i$ eats $S_j$ or $S_j$ eats $S_i$. 

However, the set of tournament digraphs with an odd number $N_S$ of vertices is a subset of a wider class of graphs known as the {\em k-regular circulant} digraphs, in which every vertex in the graph has an in-degree (number of directed edges pointing into it) of $k$ and an out-degree (number of directed edges pointing out from it) also of $k$: in RPS, each vertex in the dominance network has in-degree and out-degree of one, so the RPS dominance network graph is {\em 1-regular}; and in RPSLS each vertex has in-degree and out-degree of two, so its dominance network is {\em 2-regular}. Crucially, unlike tournament digraphs, vertices in circulant digraphs are not always connected to every other vertex in the network.

In these terms then, the $N_S$$=$$3$ RPS dominance network graph and the $N_S$$=$$5$ RPSLS dominance network graph are each {\em both} a tournament graph {\em and} a circulant digraph:  but once we move up to $N_S$$=$$7$ it is possible to study the behavior of RPSLS-like systems where the dominance network is {\em either} a tournament network (every species directly interacting with all others) {\em or} a non-tournament circulant (NTC) network (each species interacting with some subset of $2k$$<$$N_S$ other species in the system). Other authors (e.g.\ \cite{yang_park_2023}) have reported studies of 7-species extensions of RPSLS, but always as tournament systems. As far as I am aware, this is the first study of RPSLS for $N_S$$>$$5$ via NTC dominance networks. 

Here I explore what Zhong et al.\ called ecosystem interaction structures, i.e.\ dominance networks, that are a subset of the $n$-node circulant digraphs $D(N_S,\Omega)$ for odd-numbered $N_S$ and $|\Omega|$$=$$2$. In the terminology of circulants, the 3-vertex RPS game dominance network can be seen as the 1-regular circulant digraph $D(3,\{1\})$ (it has three vertices, and each vertex $v_i$  dominates vertex $v_{i\oplus 1}$ where $\oplus $ denotes addition modulo $N_S$) and the 5-vertex RPSLS dominance net can be seen as  the 2-regular circulant $D(5,\{1,3\})$ (it has five vertices, and each vertex $v_i$ dominates vertices $v_{i \oplus 1}$ and $v_{i \oplus 3}$). This observation then allows us to generalise, creating RPSLS-like systems with higher values of odd $N_S$ by using the set of distinct digraphs $D(N_S,\{1,3\})$ and/or $D(N_S,\{1,N_S\ominus2\})$ (where $\ominus$ denotes subtraction modulo $N_S$) which each have a total of $2N_S$ edges and in which each species is still predator to two species and prey to two other species. Hence the number of edges in these multi-RPSLS systems can be much larger than 10, and then deleting one edge from the digraph represents a less-than-10\% reduction in the number of species interactions. 

Having recognised that the set of  tournament digraphs are a subset of the circulants, we can in principle study ESCGs where there are (for example) 101 species:
and yet where each species remains in an RPSLS-style interaction, being predator to only two other species, and being predated upon by only two other species: given that the original 5-species version of RPSLS is $D(5,\{1,3\})$,  
we would  configure the dominance network as $D(101,\{1,3\})$ or $D(101,\{1,99\})$. This insight opens up the possibility of exploring any changes in dynamics of the class of $D(N_S,\Omega)$ systems for $|\Omega|$$\ll$$N_S$ as $N_S$ is increased up from $7$, which is one obvious line of future enquiry. Another route of enquiry would fix $N_S$ at some suitably high value and then explore variations in $k$$=$$|\Omega|$. But, in this paper, I start small and present a detailed examination of the 2-regular seven-species case, i.e. $k$$=$$2$ and $N_S$$=$$7$. As will be shown below, results from $N_S$$=$$7$ cast unexpected doubts on the generality of the key result reported by \cite{zhong_etal_2022_ablatedRPSLS}.

The novel contributions of this paper are the recognition of RPS and RPSLS dominance networks as $k$-regular circulant digraphs, and the presentation of results from extensive computer simulation studies which show that the central result of \cite{zhong_etal_2022_ablatedRPSLS}, their observed collapse in biodiversity down to two or fewer species, is due to the specific fully-connected tournament dominance network that they used in their RPSLS model: when other, equally realistic, RPSLS-style dominance networks are used instead, no such collapse in biodiversity occurs.      

Section~\ref{sec:background} explains the background to this work, and then Section~\ref{sec:graphtheory} briefly reviews the relevant graph theory needed to understand the distinction between tournament and circulant digraphs. After that, Section~\ref{sec:DoE} explains the design of experiments used here, and Section~\ref{sec:results} then shows results both from many hundreds of individual simulations. The results are discussed in Section~\ref{sec:discussion}.

\section{Background}
\label{sec:background}

\subsection{Three Species: Rock, Paper, Scissors}

\begin{figure}
\begin{center}
\includegraphics[trim=0cm 0cm 0cm 0cm, clip=true,scale=0.4]{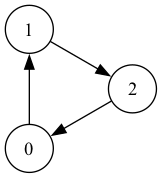}
\end{center}
\caption{The dominance network for the Rock-Paper-Scissors (RPS) game visualised as a three-vertex directed graph. A directed edge from vertex $v_i$ to vertex $v_j$ denotes that the move represented by $v_i$ dominates the move represented by $v_j$. There are three possible distinct labelings of the vertices: $(v_0$$=$$R$, $v_1$$=$$S$, $v_2$$=$$P)$; $(v_0$$=$$P$, $v_1$$=$$R$, $v_2$$=$$S)$; and $(v_0$$=$$S$, $v_1$$=$$P$, $v_2$$=$$R)$. The three distinct graphs are {\em isomorphic}, i.e. their topology is identical and they differ only by their labelings.}
\label{fig:RPS_net}
\end{figure}

In its original form, Rock-Paper-Scissors (RPS) is a hand-gesture game played by two competitors who both, at the same instant in time, make their move by forming one of their hands into either a fist ({\em Rock}), a flat open palm ({\em Paper}), or by making a vee-shape with their index and middle fingers ({\em Scissors}). If the two players have each made the same move, the game is tied; but if the two gestures are different then the winner is determined according to the three rules that define the game: {\em Rock} beats (``blunts'') {\em Scissors}; {\em Scissors} beats (``cuts'') {\em Paper}; and {\em Paper} beats (``covers'') {\em Rock}. In the language of mathematical game theory, RPS is a {\em zero-sum game} because the winner's gain and the loser's loss are equal: that is, they sum to zero.  Furthermore, and using  using R, P, and S as abbreviations for the three possible moves, game theorists would say that R {\em dominates} S; S dominates P; and P dominates R -- and this {\em intransitive dominance network} means that RPS is often referred to as a {\em cyclic} game, because when the dominance network is visualised as a three-vertex directed graph or {\em digraph} (as shown in Figure~\ref{fig:RPS_net}) the entire network is a single cycle or loop. For a game to be transitive, if move $X$ beats move $Y$ and move $Y$ beats move $Z$ then move $X$ would also beat move $Z$, and the dominance network would be a directed acyclic graph.

Despite its stark simplicity, the RPS game has proven to be a remarkably useful tool in theoretical biology where the complexities of the intricate and dynamically changing network of competitive interactions between species in an ecosystem has often been modelled by reducing the inter-species interactions to the playing of simple games such as the three-move RPS game or the equally well-known two-move game {\em  Prisoners Dilemma} (see e.g.\ \cite{nowak_may_1992,huberman_glance_1993}), well known not only for its use in modelling biological ecosystems but also for its value as a model and a metaphor in the social, economic, and political sciences.

Landmark papers exploring RPS-based models of ecosystems were published in 2007--08 by \cite{reichenbach_mobilia_frey_2007_nature,reichenbach_mobilia_frey_2007_physrevlet,reichenbach_mobilia_frey_2008_jtb}, who extended the previous non-spatial RPS model of \cite{may_leonard_1975}  by instead modeling each species as a time-varying number of discrete individual agents, where at any one time each agent occupies a particular cell in a regular rectangular lattice or grid of cells, and can {\em move} from cell to cell over time --- that is, the agents are {\em spatially located} and {\em mobile}. Agents can also, under the right circumstances, {\em reproduce} (asexually, cloning a fresh agent of the same species into an adjacent empty cell on the lattice); and they can also {\em compete} with an individual agent in a neighbouring cell by playing the RPS game, with the loser being deleted from the lattice leaving an empty cell behind. Different authors use different phrasings to explain the inter-species competition, but it is common to talk in terms of predator-prey dynamics: that is, each species is predator to (i.e., {\em dominates}) some specified set of other species, and is in turn also prey to (i.e., is {\em dominated} by) some set of other species. Because this class of models involves multiple species {\em evolving} over time while being {\em spatially} located and mobile, and competing via the {\em cyclic} dominance network of a simple {\em game}, they are commonly referred to as {\em Evolutionary Spatial Cyclic Game} (ESCG) models. 


\subsection{Five Species: Rock, Paper, Scissors, Lizard, Spock}

\begin{figure} 
\begin{center}
\includegraphics[trim=0cm 0cm 0cm 0cm, clip=true, scale=0.3]{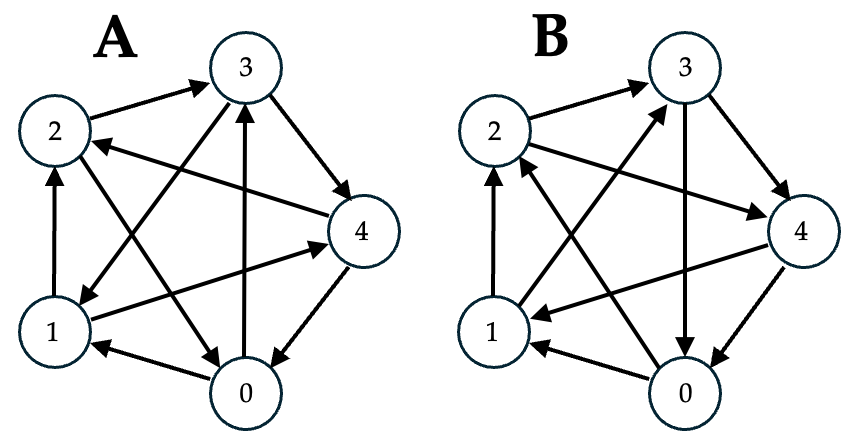}
\end{center}
\caption{Two five-species dominance networks. Network $A$ (left) is the dominance network for the Rock-Paper-Scissors-Lizard-Spock (RPSLS) game presented as a five-vertex directed graph. As with Figure~\ref{fig:RPS_net}, a directed edge from vertex $v_i$ to vertex $v_j$ indicates that $v_i$ beats $v_j$. One of the five isomorphic labelings of this graph is  
$v_0$$=$$S$ (Scissors), 
$v_1$$=$$P$,
$v_2$$=$$R$, 
$v_3$$=$$L$ ,  and 
$v_4$$=$$V$ (for Vulcan, the planet Mr Spock was born on). The rules of this game are: scissors cut paper;
paper covers rock; 
rock blunts scissors;
scissors decapitates lizard;
lizard eats paper;
paper disproves Spock;
Spock vaporizes rock;
rock crushes lizard;
lizard poisons Spock; and
Spock smashes scissors. In the graph-theory terminology introduced in Section~\ref{sec:graphtheory}, this network is the 2-regular tournament circulant digraph $D(5,\{1,3\})$. The Eulerian circuit visiting all the Network~A nodes using only the edges forming the convex hull of the outer pentagon traces a clockwise cycle, and the Eulerian circuit traversing only the edges forming the inner pentangle traces a counterclockwise route. 
Network~$B$ (right) is the tournament circulant digraph $D(5,\{1,2\})$: this is superficially different from Network~$A$ because its inner-pentangle circuit is clockwise rather than counterclockwise; but $B$ is isomorphic with $A$ (denoted by $A$$\cong$$B$) under the relabelling $0_B$$\rightarrow$$0_A$, $1_B$$\rightarrow$$3_A$, $2_B$$\rightarrow$$1_A$, $3_B$$\rightarrow$$4_A$, $4_B$$\rightarrow$$2_A$. 
 }
\label{fig:RPSLS_net}
\end{figure}

\cite{kass_bryla_1998} invented an extension of RPS, adding two more moves inspired by the classic science-fiction TV and movie franchise {\em Star Trek}: these are {\em Lizard} and {\em Spock}, and the expanded game is most often referred to by the acronym RPSLS. In {\em Star Trek}, Mr Spock is from the planet Vulcan, and to avoid the clash with S already abbreviating {\em Scissors}, the character V can be used as the single-letter abbreviation for {\em Spock} in the RPSLS dominance network, which is illustrated in Figure~\ref{fig:RPSLS_net}.

RPSLS can be played by people as a hand-gesture game in the same style as the original RPS: {\em Lizard} has a gesture best described as a naked sock-puppet; while {\em Spock}'s gesture involves the hand being held vertical, open palm facing the other player and with the fingers held such that there is a vee-shaped separation between the middle finger and the ring finger (this is a gesture the {\em Star Trek} character uses). The caption to Figure~\ref{fig:RPSLS_net} explains the dominance relations, i.e.\ which move beats which other moves.

Simulations of co-evolutionary population dynamics via ESCGs are simple discrete-time systems that are technically unchallenging to write a program for, and are strongly reminiscent of -- but not identical to -- cellular automata (see e.g. \cite{wolfram_2002_book}). The lattice/grid needs to first be set up, i.e. its dimensions and initial conditions at the first time-step need to be specified. Each cell in the grid is either empty or contains exactly one agent, and each individual agent is a member of exactly one of the model's set of species. If the number of species in the model is $N_s$, one common way of doing the initialisation is to assign one individual to every cell in the grid, with that individual's species being an equiprobable choice from the set of available species (i.e., choose species $s_i$ with probability $1/N_s$). The modeller also needs to specify the dimensionality of the lattice, and its {\em length} (i.e., number of cells) along each dimension. In almost all of the published work in this field, the lattice is two-dimensional and square, so its extent is defined by a single system parameter: the side-length (conventionally denoted by $L$); and the total number of cells in the lattice (conventionally denoted by $N$) is hence $N=L^{2}$. Working with 2D lattices has the advantage that the global state of the system can be readily visualised as a snapshot at time $t$ as a color-coded or gray-shaded 2D image, with each species in the model assigned its own specific color or gray-scale value, and animations can easily be produced visualising the change in the system state over time. 

In the literature on evolutionary spatial cyclic games, authors often make the distinction between two scales of time-step in the simulation. At the very core of the simulation process is a loop that iterates over a number of {\em elementary steps} (ESs), the finest grain of time-step; and then some large number of consecutive ESs is counted as what is conventionally referred to as a {\em Monte Carlo Step}.  

In a single ES, one individual cell (denoted $c_i$) is chosen at random, and then one of its immediately neighboring cells (denoted $c_n$) is also chosen at random: most published work on 2D lattice model ESCGs uses as the set of possible neighbours the 4-connected von Neumann neighborhood rather than the 8-connected Moore neighborhood commonly used in cellular automata research; but there seems to be no firm convention on whether to use periodic boundary conditions (also known as toroidal wrap-around) or no-flux boundary conditions (such that cells at the edges and corners of the lattice have a correspondingly reduced neighbour-count) -- some authors use periodic, others no-flux. The results presented in this paper come from simulations with no-flux boundary conditions, as used by \cite{zhong_etal_2022_ablatedRPSLS}. Then in each ES one or more of three possible actions occurs: competition, reproduction, or movement, and the probabilities of each of these three actions occurring per ES is set by system parameters $\mu, \sigma,$ and $\epsilon$, respectively (this is explained in more precise detail later, in Section~\ref{sec:background}). {\em Competition} involves the individuals at $c_i$ and $c_n$ interacting according to the rules of the cyclic game, resulting in either a draw or one of the individuals losing, in which case it is deleted from its cell, leaving an empty cell (denoted by $\emptyset$); {\em reproduction} occurs when one of $c_i$ or $c_n$ is empty, the empty cell being filled by a new individual of the same species as the nonempty neighhbor; and {\em movement} involves swapping the contents of $c_i$ and $c_n$.  

In the original formulation, each ES involves only one of the three possible actions (competition, reproduction, or movement) occurring for a single cell, and a Monte Carlo Step (MCS) is conventionally defined as a sequence of $N=L^2$ ESs, the rationale being that, on the average, each cell in the lattice will be randomly chosen once per MCS, and hence that, again on the average, every cell in the grid has the potential to change between any two successive MCSs. Most published research on this type of model uses MCS as the unit of time when plotting time-series graphs illustrating the temporal evolution of the system, and I follow that convention here. 

In their seminal papers, \cite{ reichenbach_mobilia_frey_2007_nature,reichenbach_mobilia_frey_2007_physrevlet,reichenbach_mobilia_frey_2008_jtb} studied 2D lattice systems where interspecies competition was via $N_S$$=$$3$ RPS games, with $\mu$$=$$\sigma$$=$$1.0$, and where $L$ ranged from 100  to 500, and they illustrated and explained how the overall system dynamics result in progressive emergence of one or more temporally and spatially coherent interlocked {\em spiral waves}.
The specific nature of the wave-patterning, i.e. the size and number of spiral waves seen in the system-snapshot images, depended on a {\em mobility} measure $M=\epsilon/2N$, which is proportional to the expected area explored by a single mobile individual in the model, per MCS.

In the years since publication of \cite{reichenbach_mobilia_frey_2007_nature,reichenbach_mobilia_frey_2007_physrevlet,reichenbach_mobilia_frey_2008_jtb}, many papers have been published that explore the dynamics of such ESCGs based on RPS. 
For examples of recent publications exploring a range of issues in the three-species RPS ESCG, see: \cite{
nagatani_ichinose_tainaka_2018_RPS,
kabir_tanimoto_2021_RPS,
mohd_park_2021_RPS,
bazeia_bongestab_deoliveira_2022_RPS,
park_2021_RPS,
menezes_batista_rangel_2022_RPS,
menaces_rangel_moura_2022_RPS,
zhang_bearup_guo_zhang_liao_2022_RPS,
menezes_barbalho_2023_RPS,park_jang_2023_RPS} and \cite{kubyana_landi_hui_2024_RPS}.

More recently, various authors have reported experiments with a closely related system where $N_S$$=$$5$: this game is known as Rock-Paper-Scissors-Lizard-Spock (RPSLS), an extension of RPS introduced by \cite{kass_bryla_1998} and subsequently publicized in an episode of the popular American TV show {\em Big Bang Theory}. The dominance network for the RPSLS game is illustrated in Figure~\ref{fig:RPSLS_net} and explained in the caption to that figure.  This (and other five-species ECSGs) was first explored in the theoretical biology literature by \cite{laird_schamp_2006_RPSLS,laird_schamp_2008_RPSLS,laird_schamp_2009_coexistence_RPSLS}, and other recent papers exploring RPSLS include 
\cite{hawick_2011,
kang_pan_wang_he_2013,
cheng_yao_huang_park_do_lai_2014,
park_do_jang_lai_2017,
park_jang_2019,
park_jang_2020} 
and \cite{viswanathan_etal_2024}; 
and ECSGs related to RPSLS, but with $N_S$$\geq$$5$, were recently explored by \cite{avelino_deoliveria_trintin_2022_RPS_bigN}.

\subsection{Zhong et al.\ (2022): Ablated Dominance Networks}

In a recent paper, \cite{zhong_etal_2022_ablatedRPSLS} studied the effects on co-evolutionary dynamics of selectively ablating the RPSLS dominance network, i.e., deleting one or more of the directed edges in the RPSLS digraph and exploring the consequent changes in the population dynamics. In the abstract to their paper, Zhong et al.\ wrote that these systematic changes to the dominance network (which they refer to as the {\em interaction structure}):
\begin{quote}
 ``$\ldots$ impacts the evolutionary dynamics, and different interaction structures allow for different states of multi-species coexistence. We also find that the competition between different three-species-cyclic interactions is crucial for the realization of different asymptotic behaviors at low mobility. Our findings may be useful to understand the subtle effects of competitive structure on species coexistence and evolutionary game outcomes."  
\end{quote}

Here I will use $N_a$ to denote the number of ablated edges in the dominance network. Specifically, Zhong et al.\ show (in their Figures~3, 5, 6, and~7) results from many thousands of independent repetitions of simulations of the ablated-digraph RPSLS systems for networks with $N_a \in \{1, 2, 3\}$ ablated edges, plotting the frequencies of different classes of outcome for mobility $M$ values ranging from $10^{-7}$ to $10^{-3}$. Zhong et al.'s primary measure of the outcome of any one experiment was the number of species remaining in what they refer to as the ``asymptotic state'', i.e.\ how many species remain with nonzero population-counts at time $t$$=$$10^{5}$MCS -- the assumption being that by the time this many MCS have been simulated, the system will have settled to a steady-state dynamic equilibrium.   Here I will use $n_s(t)$ to denote the number of species remaining after $t$ MCS, so Zhong et al.'s key metric is $n_s(10^5)$. Using this notation, Zhong et al.'s  primary observation is that for each value of $N_a$,  at low $M$, the system almost always converges to $n_s(10^5)$$\in$$\{3,4\}$, and then as $M$ is increased the system shows a gradual fall to zero in the frequency of $n_s(10^5)$$=$$4$, followed by a sudden steep decline in frequency of $n_s(10^5)$$=$$3$, with $n_s(10^5)$$\leq$$2$ becoming the asymptotic state in $\approx$$100\%$ of the experiments. In all of Zhong et al.'s  experiments, this sudden transition, the change from $n_s(10^5)$$=$$3$  to $n_s(10^5)$$\leq$$3$, occurs very sharply somewhere in the interval $M$$\in$$[10^{-4}/2,10^{-4}]$. 

Zhong et al.\ used a square grid with L=200 and ran each experiment for $10^{5}$ MCS. For each data-point on their asymptotic-state frequency plots they computed 500 independent and identically distributed (IID) repetitions of any one experiment for any given value of $M$, and each of their Figures~3 and~5--7 has data-point markers showing that they sampled 20 different values of $M$$\in$$[10^{-7},10^{-3}]$ for each plot. Thus, in total, the four asymptotic-state frequency figures presented by Zhong et al.\ represent results from $4 \times 20 \times 500 \times 10^{5} = 4.0$$\times$$10^9$ individual MCS, and each of their MCS involves $L^2$$=$$4$$\times$$10^5$ ESs, so the total number of ESs they simulated is $1.6$$\times$$10^{14}$. That's 160 trillion ESs in total.

\subsection{Summary}
To summarise: for RPS and RPSLS the {\em only} viable dominance network is a tournament, because the values of $N_S$ are so low. When higher values of $N_S$ are used to create the circulant dominance networks, a richer space of topologies opens up, offering the possibility of modelling $N_S$-species ecosystems coexistence and biodiversity in situations where it is {\em not}\/ the case that every single species interacts (either positively or negatively) with every other species in the entire ecosystem, which must surely be the case in many real-world situations of genuine interest. Yet, to the best of my knowledge, it seems that every researcher working on RPS-style ESCGs focuses either on games defined by fully-connected tournament networks, or on games where the dominance network is a simple single macrocycle where all nodes have in-degree and out-degree of one. As far as I am aware, my work reported here is the first to explore RPS-style ESCGs with non-tournament circulant graphs where the node in-degrees and out-degrees are greater than one.

\section{Graph Theory for Dominance Networks}
\label{sec:graphtheory}
\label{sec:graphtheory}
 
In the language of graph theory, we can say that the RPS and RPSLS dominance networks shown in Figures~\ref{fig:RPS_net} and~\ref{fig:RPSLS_net} are both {\em strongly connected} digraphs, because in both for each vertex all other vertices in the graph are reachable by at least one path; they are also both {\em regular} digraphs, because in both each node/vertex has the same {\em in-degree} (number of arrows pointing into it) and the same  {\em out-degree} (number of edges pointing out of it) and the in-degree and out-degree are equal (a digraph is said to be {\em k-regular} if each vertex has in-degree and out-degree of $k$, so the RPSLS network is 2-regular while the RPS network is 1-regular). More specifically, these two graphs are instances of what is known as an $n$-node  {\em tournament}. An $n$-node tournament graph is formed by starting with the undirected $n$-node {\em complete graph} (i.e., where each node/vertex is directly connected to all $n$$-$$1$ other nodes in the graph) and then giving the graph an {\em orientation}, which involves turning each undirected edge in the graph into a directed edge. 

The number of edges in an $n$-node complete graph is given by $n(n-1)/2$, and hence that is also the number of directed edges in a tournament digraph. Tournament digraphs are useful to us here because there is only a single directed edge between any two vertices: this is important in the present context because domination, i.e.\ which move beats the other for any pair of moves, is absolute and mutually exclusive: either node $A$  dominates node $B$, represented by an arrow from $A$ to $B$, or $B$ dominates $A$, represented by an arrow from $B$ to $A$; there would be little sense in a dominance network where there is an arrow from $A$ to $B$ and {\em also} an arrow from $B$ to $A$, that would be a contradiction -- below, when discussing procedurally constructed circulant digraphs, I'll refer to this as the {\em unidirectionality constraint}. 

Some further terminology will be useful in the discussion that follows: if one or more directed paths exists between two nodes in a digraph then the shortest path between them is is known as the {\em graph geodesic} and the number of edges in the geodesic is the {\em distance} between the two vertices. The {\em diameter} of a graph, denoted here by \diameter, is the length of the longest geodesic in the graph (this can be identified by creating the set  $S$ of shortest paths for each pair of vertices in the network and then identifying the longest path in $S$: that is, the diameter is the longest shortest path). 

Further, note that because we want  dominance network digraph to be regular (so that each species is dominated by the same number of species that it dominates, as in RPS and RPSLS), for a tournament network the count of $n-1$ edges connecting to each node must be even (half coming in, half going out), and hence the total number of nodes in the network $n$ must be an odd number. And, finally, using the definition given by \cite{vandoorn_1986}, a {\em circulant digraph} (also known as a {\em directed star polygon})  denoted by $D(n,\Omega)$ consists of a set of vertices $V(D) = \{v_0, v_1, \ldots, v_{n-1}\}$ and a set of arcs (directed edges) $A(D)$ such that:
\[   (v_i, v_j) \in A(D) \iff (j-i) \equiv \omega \pmod n \] 
for some $\omega \in \Omega$ where $\Omega \subset \{1, 2, \ldots, n-1\}$. In these terms, the RPS dominance network is $D(3,\{1\})$ and the RPSLS network is $D(5,\{1,3\})$. In the discussion that follows, it will be useful consider the set of vertices as an ordered sequence, i.e.\ $V(D)=\langle v_0, v_1, \ldots, v_{n-1}\rangle$; and let the symbols $\oplus$ and $\ominus$ denote addition and subtraction modulo $n$ for vertex-index arithmetic, such that for example: if $i$$=$$0$ then
$ v_{i \ominus 1} =    v_{n-1},$  otherwise $v_{i \ominus 1} = v_{i-1}$; 
and if $i$$=$$n$$-$$1$ then 
$ v_{i \oplus 1} = v_{0},$ otherwise $v_{i \oplus 1}= v_{i+1}$.

\begin{figure}
\begin{center}
\includegraphics[trim=0cm 0cm 0cm 0cm, clip=true,scale=0.3]{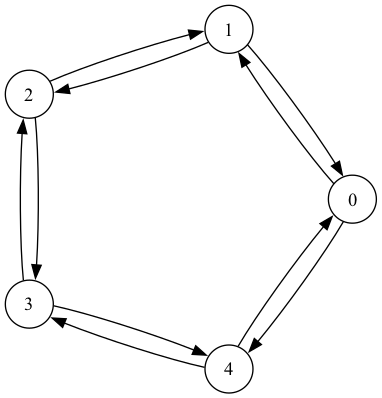}
\includegraphics[trim=0cm 0cm 0cm 0cm, clip=true,scale=0.3]{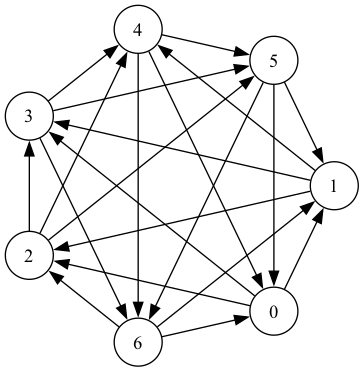}
\includegraphics[trim=0cm 0cm 0cm 0cm, clip=true,scale=0.3]{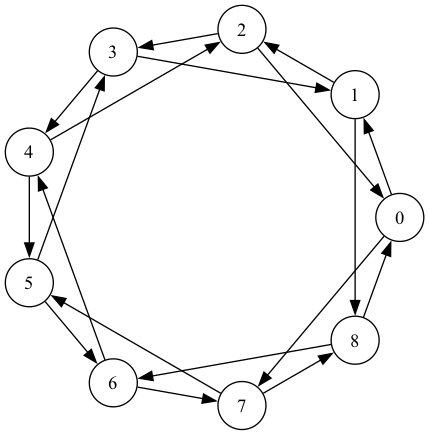}
\end{center}
\caption{Three examples of circulant digraphs: upper-left network is $D(5,\{1, 4\})$, which has a diameter \diameter$=$$2$, and which violates the unidirectionality constraint introduced in Section~\ref{sec:graphtheory}; upper-right is the {\em tournament} circulant digraph $D(7,\{1, 2, 3\})$, with \diameter$=$$2$; below is a ``generalised RPSLS'' non-tournament circulant (NTC) digraph $D(9,\{1, 7\})$, with \diameter$=$$4$.}
\label{fig:circulants}
\end{figure}

Procedurally, a graph of an $N_S$-vertex circulant can be constructed by placing the $n$ vertices equally spaced around the circumference of a circle, and then for each vertex $v_i$ doing the following: for each $\omega_k \in \Omega$ draw a directed arc from $v_i$ to $v_j$ where $j = i\oplus k $. That is, the elements of $\Omega$ are a set of vertex-index-offsets to be used for connecting each  $v_i$ to the other nodes that it dominates; repeating this for all nodes creates the full network.  Examples of circulant digraphs are illustrated in Figure~\ref{fig:circulants}. Because of the unidirectionality constraint, the need for the dominance networks to be regular, and the requirement that $n$ must be odd, we need only concern ourselves with those circulants where $|\Omega|$$\leq$$\lfloor$$N_S/2\rfloor$. 
An additional constraint to avoid redundancy is to let $\Omega$$=$$\{\omega_1, \omega_2, \ldots, \omega_{|\Omega|}\}$ and require 
$\omega_i$$<$$\omega_{i+1},\forall i$. 
When $|\Omega|$$=$$\lfloor$$N_S/2\rfloor$ a tournament graph is produced but, as is shown in Figure~\ref{fig:circulants}, for $N_S$$>$$5$ interesting topologies can be generated by cases where $|\Omega|$$<$$\lfloor$$N_S/2\rfloor$. 

Here I'll assert that the combinatorics of $\Omega$ only start to get properly interesting once $N_S$ rises above the $5$ of RPSLS. To explain that claim, first consider this: when generating an $N_S$$=$$3$ circulant for RPS, there are only two possible contents of $\Omega$: $\{1\}$ and $\{2\}$; and the two networks generated by these are {\em isomorphic} tournaments -- i.e., topologically identical apart from the (arbitrary) labelling of the vertices. So there is only one possible RPS dominance network: it is 1-regular, and is {\em both} a tournament digraph {\em and} a circulant digraph: D(3,\{1\}).

Similarly, the RPSLS dominance network of Figure~\ref{fig:RPSLS_net}A is 2-regular, and also it is {\em both} a tournament digraph {\em and} a circulant digraph: $D(5,\{1,3\})$, and it is unique. There is another $D(5,S)$$:|S|$$=$$2$ graph that is superficially different from $D(5,\{1,3\})$, and that is the circulant $D(5,\{1,2\})$, illustrated in Figure~\ref{fig:RPSLS_net}B. Comparing Figures~\ref{fig:RPSLS_net}A and~\ref{fig:RPSLS_net}B, the cyclic path $v_0$$\rightarrow$$v_1$$\rightarrow$$v_2$$\rightarrow$$v_3$$\rightarrow$$v_4$$\rightarrow$$v_0$ traced around the outer pentagon is clockwise in both graphs; but the cyclic path on the inner pentagram is anticlockwise $v_0$$\rightarrow$$v_3$$\rightarrow$$v_1$$\rightarrow$$v_4$$\rightarrow$$v_2$$\rightarrow$$v_0$ in the RPSLS dominance net of $D(5,\{1,3\})$ but is clockwise $v_0\rightarrow$$v_2$$\rightarrow$$v_4$$\rightarrow$$v_1$$\rightarrow$$v_3$$\rightarrow$$v_0$ in the alternate dominance net of $D(5,\{1,2\})$ -- the nodes of which could just as easily be given labels of {\em Rock}, {\em Paper}, {\em Scissors}, {\em Lizard}, and {\em Spock}.  The caption of Figure~\ref{fig:RPSLS_net} explains how the vertices of $D(5,\{1,2\})$ can be relabelled to give $D(5,\{1,3\})$ and hence $D(5,\{1,2\}) \cong D(5,\{1,3\})$. Thus, like the $N_S$$=$$3$ case of RPS, there is only one, unique, dominance network, which is  {\em both} a tournament digraph {\em and} a 2-regular circulant digraph. 

But, as I show below, when $N_S$$>$$5$, (in fact, as soon as $N_S$$=$$7$), there are $k$-regular RPSLS-style dominance networks that are tournament circulant digraphs like those for $N_S$$\in$$\{3,5\}$, but there are {\em also} $k$-regular RPSLS-style dominance networks that are circulant digraphs that are {\em not} tournaments -- and these {\em non-tournament circulant }(NTC) digraphs  are of significant interest. For $N_S$$=$$7$, the tournament circulant digraphs are 3-regular, but the NTC networks of relevance for this paper are 2-regular, because \cite{zhong_etal_2022_ablatedRPSLS} reported results from the original RPSLS, and the original RPSLS is 2-regular.

Before going on to explore NTC digraphs in more detail, a final issue to discuss is how to determine the number of distinct (non-isomorphic) 2-regular circulant digraphs for $N_S$$\geq$$5$.  

Consider first the $N_S$$=$$5$ case: when generating circulant for RPSLS the contents of $\Omega$ satisfying $|\Omega| \leq \lfloor N_S/2\rfloor$ can in principle be any of the following: $\{1\}$; $\{2\}$; $\{3\}$; $\{4\}$; $\{1, 2\}$; $\{1, 3\}$; $\{1, 4\}$; $\{2, 3\}$; $\{2, 4\}$; and $\{3, 4\}$. 
The networks given by $D(5,\{1, 4\})$ (illustrated in Figure~\ref{fig:circulants}) and $D(5,\{2, 3\})$ both violate the unidirectionality constraint because for any 2-regular circulant where $\Omega$$=$$\{\omega, N_S \ominus \omega \}$, each vertex $v_i$ will dominate $v_{i\oplus \omega}$ and each vertex $v_{i\oplus \omega}$ will dominate $v_i$; that is, here  $\{1, 4\} \equiv \{1, n \ominus 1\}$ and $\{2, 3\} \equiv \{2, n \ominus 2\}$ and hence those two values of $\Omega$ are ruled out of consideration. 
The four cases where $|\Omega|$$=$$1$ give four isomorphic networks (i.e., only one distinct graph) that topologically is a simple 1-regular $N_S$-node loop: evolutionary spatial cyclic games (ESCGs) involving such loops have been explored in previous publications (see e.g.\ \cite{carvalho_mota_martins_2023}), but are not of interest here because they are not 2-regular -- they can be considered as $N_S$$>$$3$-node extensions of RPS, but are not analogous to RPSLS. Note that, because of the modular index-arithmetic of ordered vertices on 2-regular $N_S$-node circulant graphs, a pair of offsets $\Omega$$=$$\{\omega_1, \omega_2\}$ is equivalent to the pair of offsets $\Omega$$=$$\{N_S\ominus \omega_1,N_S \ominus \omega_2\}$. So of the four remaining $| \Omega |$$=$$2$ cases for $N_S$$=$$5$, $D(5,\{1, 2\}) \cong D(5, \{3, 4\})$ (that is, they produce isomorphic graphs) and  $D(5,\{1, 3\}) \cong D(5, \{2, 4\})$.  This leaves $\Omega$$=$$\{1, 2\}$ and  $\Omega$$=$$\{1, 3\}$, which produce a pair of isomorphic graphs, as explained in the caption to Figure~\ref{fig:RPSLS_net}.

The reasoning for $N_S$$=$$5$ given in the previous paragraph can be generalised for 2-regular circulants with $N_S$$>$$5$ in very few words, as follows: we need the set of all offset pairs 
$ \Omega_i$$=$$\{\omega_{i,1}, \omega_{i,2}\}$$\in$$\{1, \ldots, N_S-1\}^2 $
such that
$ \omega_{i,2}$$>$$\omega_{i,1}$ 
and 
$ \omega_{i,2}$$\neq$$N_S$$\ominus$$\omega_{i,1}$. 

For further reading on the mathematics of regular tournaments, see \cite{chamberland_herman_2014,akin_2020} and \cite{akin_2023}; on generating regular digraphs see \cite{brinkmann_2013}.

\section{Design of Experiments}
\label{sec:DoE}

Section~\ref{sec:results} presents first results from a series of seven-species ($N_S$$=$$7$) evolutionary spatial cyclic game (ESCG) simulation experiments designed to identify and explore the differences, if any, between ECSGs with tournament and circulant dominance networks, when the dominance network is subject to random ablations in the manner studied by \cite{zhong_etal_2022_ablatedRPSLS}. In all the experiments presented here, I followed the convention of setting (without loss of generality) $\mu$$=$$\sigma$$=$$1.0$ and then varying the mobility $M$ (which determines $\epsilon$) over several orders of magnitude. For each value of $M$ explored, I ran $N_{\text{exp}}$$=$$100$ independent and identically distributed (IID) ECSG simulations. The unit of time $t$ in each simulation experiment here is a single Monte Carlo Step (MCS), and all the experiments reported here ran from $t$$=$$0$ to  $t$$=$$2$$\times$$10^{5}$ MCS. In all the experiments reported here, I used a square lattice with side-length $L$$=$$500$, and the lattice was initialised by each cell being filled with an agent whose species was set by random draw from a uniform distribution such that the probability of the individual agent $i$'s species-type $s_i$ being any specific species $S_j$ was ${\text Pr}(s_i=S_j)=\frac{1}{7}, \forall i, \forall j$.  

The set of nested control loops for the main experiment is shown as pseudocode in  Algorithm~\ref{alg:main_exp}. 
This takes as control parameters the value of $N_S$, the set of offsets $\Omega$ that determined the dominance network connectivity, and the number of directed edges in the network to be randomly ablated. It first creates the full unablated network, and then ablates the specified number of randomly-chosen edges in the network. In all the experiments reported here, only one randomly-chosen edge was ablated, and this was always ablated from the set of edges originating from species $S_0$: that is, in each ablation experiment, the number of other species dominated by $S_0$ is one fewer than the number dominated by all other species. 

Following \cite{zhong_etal_2022_ablatedRPSLS}, the primary outcome that was monitored for each individual experiment was the number of species remaining at the final MCS, denoted here by $n_s(T_{\text{max}})$ but whereas Zhong et al.\ used  $n_s(10^5)$, the experiments reported here were run for considerably longer than that, because the dynamics of the 7-species systems were observed to have much longer time constants than those of the 5-species system explored by Zhong et al.: whereas the $N_S$$=$$5$ systems studied by Zhong et al.\ could stabilise to an ``asymptotic state'' by $T_{\text{max}}$$=$$10^5$, the $N_S$$=$$7$ systems studied here typically needed $T_{\text{max}}$$\geq$$2$$\times$$10^5~$ before they settled to a stable state.

In each individual ECSG experiment, the density $\rho_i(t)$ of each species $S_i$ was recorded after each MCS: illustrative time-series of the $\rho_i$ values from several experiments are shown in Section~\ref{sec:results}. 
The variation in densities at any one time is also of interest: for this the mean density at time $t$ was calculated as $\hat{\rho}(t)=\frac{1}{N_S}\sum^{N_S}_1 \rho_i(t)$ and then the variation in density $\rho_v(t)$ was calculated as the standard deviation 
$\rho_v(t)=(\frac{1}{N_S}\sum^{N_S}_1 (\rho_i(t)-\hat{\rho}(t))^2)^{0.5}.$

For each IID repetition of the experiment, the dominance network is created and randomly ablated, and the lattice for time $t$$=$$0$ is populated with agents of randomly assigned species, as described above. Then for each $M$-value being explored, the lattice is reset to its $t$$=$$0$ state and then $T_{\text{max}}$ MCS are executed. Each MCS consists of executing some number (denoted here by $E_\text{max}$) of {\em elementary steps} (ESs). Before each ES, an individual cell is in the lattice is randomly selected: let $\vec{p_i}$ denote its position in the lattice (i.e., in the 2D case here, $\vec{p_i}$ will be the row and column lattice coordinates of the randomly selected cell), and then one of that cell's immediate neighbors is randomly chosen for interaction -- let $\vec{p_n}$ denote the position of the neighboring cell.  The random selection of an individual cell is achieved by a simple function {\sc RndCell} and the random choice of neighboring cell is achieved by a simple function {\sc RndNeighbor}, which needs to implement two things: (1) a {\em neighborhood function}, determining whether each cell has either four neighbours (the von Neumann neighborhood) or eight (the Moore neighbourhood); and (2) the {\em boundary conditions}, i.e.\ whether a cell at an edge or corner of the lattice has a reduced set of neighbors (i.e., {\em no-flux} boundary conditions), or instead the edges of the lattice wrap-around giving it a toroidal topology (i.e., {\em periodic} boundary conditions). In all the experiments reported here, the von Neumann neighborhood was used with no-flux boundary conditions, and the number of elementary steps per MCS was set to $E_{\text{max}}$$=$$L^2$$=N$, as is the convention in the literature.

\begin{algorithm}
\caption{M-Sweep 2D Square ECSG}
\label{alg:main_exp}
\begin{algorithmic}[1]
\Require $L \geq 1 \in {\mathbb Z}$ \Comment{Side-length of square lattice $l$}
\Require ${\cal B} \in \{ \text{`noflux'}, \text{`periodic'}\}$ \Comment{Boundary condition}
\Require ${\cal N} \in \{ \text{`vonNeumann'}, \text{`Moore'}\}$ \Comment{Nbrhood}
\Require $\mu \in [0.0, 1.0] \subset {\mathbb R}$ \Comment{Pr(compete)}
\Require $\sigma \in [0.0, 1.0] \subset {\mathbb R}$ \Comment{Pr(reproduce)}
\Require $N_s \geq 3 \in {\mathbb Z}$ \Comment{\# species}
\Require $\Omega=\{ \omega_1, \ldots \}: 0 < \omega_i  < N_s$ \Comment{Circulant offsets}
\Require $N_a \geq 0 \in {\mathbb Z}$ \Comment{\# ablations}
\Require $M_{\text{min}} \in {\mathbb R}^+$ \Comment{Minimum $M$}
\Require $M_{\text{max}} \in {\mathbb R}^+$ \Comment{Maximum $M$}
\Require $M_\text{mul} >1 \in {\mathbb R}$ \Comment{$M$ multiplier}
\Require $T_{\text{max}} \geq 1 \in {\mathbb Z}$ \Comment{\#MCS per experiment}
\Require $E_{\text{max}} \geq 1 \in {\mathbb Z}$ \Comment{\#Elementary steps per MCS}
\Require $N_{\text{exp}} > 0 \in {\mathbb Z}$ \Comment{\# IID experiments}
\Ensure $N_s = 2j + 1 ; j \in {\mathbb Z}$ \Comment{$N_s$ must be odd}
\Ensure $M_{\text{min}} < M_{\text{max}}$ 
\State $N \gets L^2$ \Comment{\# cells in lattice}
\Ensure $M_{\text{max}} \leq \frac{1}{2N}$ \Comment{s.t. $\epsilon \in [0.0,1.0] \in {\mathbb R}$ }
\State $r \gets 0$ \Comment{$r$ is current IID repetition}
\While{$r<N_{\text{exp}}$}
	\State $D \gets  ${\sc CreateDomNet}$(N_s,\Omega)$ \Comment{Create dom net}
	\State $D' \gets {\text{\sc AblateDomNet}}(D,N_a)$ \Comment{Ablate net}
	\State $l_0 \gets  {\text{\sc Populate}}(L, N_s)$ \Comment{Fill lattice with agents}
	\State $M \gets M_{\text{min}}$
	\While{$M \leq M_{\text{max}}$}\Comment{Sweep M}
		\State $\epsilon \gets 2MN$ \Comment{Pr(move)}
		\State $ l \gets l_0$ \Comment{Set lattice to initial state}
		\State $t \gets 0$ \Comment{$t$ is time, in units of MCS}
		\While{$ t \leq T_{\text{max}}$}	
			\State $e \gets 0$ \Comment{$e$ is current elementary step (ES)}
			\While{$e < E_{\text{max}}$}
				\Comment{Core inner ES loop}
				\State$\vec{p_i} \gets {\text{\sc RndCell}}(l)$	
				\State$\vec{p_n} \gets {\text{\sc RndNeighbor}}(l, \vec{p_i}, {\cal B}, {\cal N}) $
				\State$l \gets$ {\sc ElStep}$(l, D', \vec{p_i}, \vec{p_n}, \mu, \sigma, \epsilon)$
				\State $e \gets e+1$
			\EndWhile
			\State $t \gets t+1$
		\EndWhile 
		\State $M \gets M \times M_{\text{mul}} $	
	\EndWhile
	\State $r \gets r +1 $
\EndWhile
\end{algorithmic}
\end{algorithm}

Note that in each IID repetition of the core ESCG simulation, the ablated network is created once only, and similarly the lattice is randomly populated with agents once only, and the system is re-set to these initial conditions for each successive value of $M$ explored: this means that, for any one initial lattice, the system is re-playing the temporal evolution of the same initial lattice for all values of $M$ and hence any variation in final outcome is going to be due to the change in $M$ and in the different sequence of random numbers generated for each experiment. 
In all the experiments reported here, the {\em revised elementary step} (RES) described in \cite{cliff_2024_RES} was used for the {\sc ElStep} function in Algorithm~\ref{alg:main_exp}: RES is more efficient in both space and time and gives qualitatively similar population dynamics in comparison to the {\em original elementary step} (OES) used previously by very many researchers in this field (e.g. \cite{yang_park_2023,zhong_etal_2022_ablatedRPSLS,reichenbach_mobilia_frey_2007_nature,reichenbach_mobilia_frey_2007_physrevlet,reichenbach_mobilia_frey_2008_jtb}).

 
\section{Results}
\label{sec:results}

For brevity, all results for NTC networks presented here come only from $D(7,\{1,3\})$; in future publications I will report results from other $D(7,\Omega)$$:$$|\Omega|$$=$$2$ NTC networks. In all the experiments reported here, $T_{\text{max}}$$=$$2$$\times$$10^{5}$, $\mu$$=$$\sigma$$=$$1.0$, and $L$$=$$500$; and because $\epsilon$$=$$1.0$ at $M_{\text{max}(L)}$$=$$1/2L^2$$=$$2$$\times$$10^{-6}$, that is the upper bound on the range of $M$ values explored. 

I first show baseline results from unablated ($N_a$$=$$0$) seven-species ($N_S$$=$$7$) tournament systems and NTC systems in Sections~\ref{sec:unablated_T} and~\ref{sec:unablated_NTC}, respectively. After that, in Section~\ref{sec:ablated_T}, I show results from ablated ($NS=7, N_a$$=$$1$) tournament systems which qualitatively replicate and confirm Zhong et al.'s results for their $N_S$$=$$5$ tournament systems: we see a phase transition where the number of coexisting species suddenly collapses once $M$ is increased beyond a threshold value. And then finally, in Section~\ref{sec:ablated_NTC}, I show results from ablated ($N_S$$=$$7, N_a$$=$$1$) NTC systems, directly analogous to the $N_S$$=$$5$ system studied by Zhong et al., in which the collapse in species numbers frequently does not occur. These results are discussed further in Section~\ref{sec:discussion}.

\subsection{Unablated Tournament Network}
\label{sec:unablated_T}

\begin{figure}
\begin{center}
\includegraphics[trim=0cm 0cm 0.1cm 0cm, clip=true, scale=0.4]{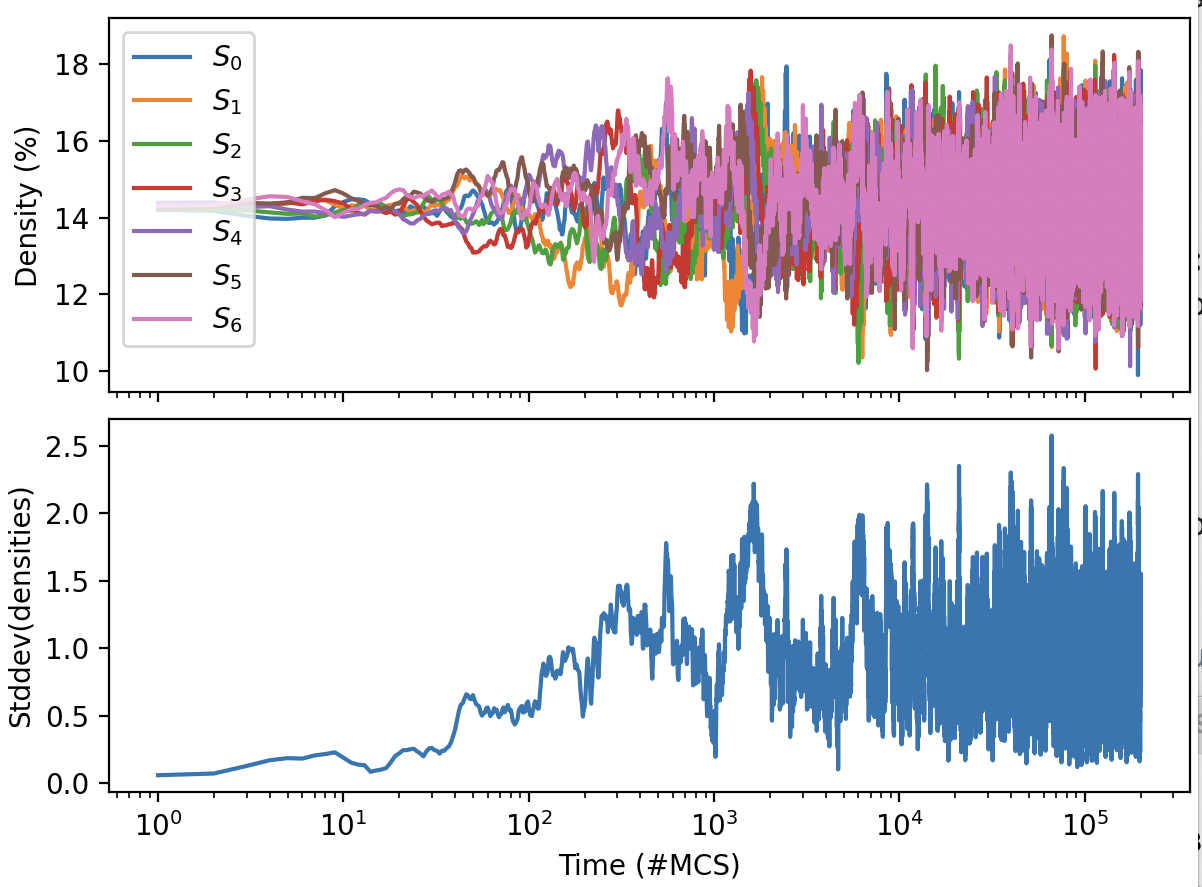}
\end{center}
\caption{{\bf Upper graph:} time-series of species densities from a single experiment with the seven-species (i.e., $N_s$$=$$7$) system when the dominance graph is the unablated (i.e., $N_a$$=$$0$) tournament circulant $D(7,\{1,3,5\})$, for $L$$=$$500$, $\mu$$=$$\sigma$$=$$1.0$, $M$$=$$10^{-7}$. Horizontal axis is time $t$ measured in MCS; vertical axis is percentage density (i.e., $\rho_i(t)$) for species $S_i$$:$$i$$\in$$\{0, \ldots, 6\}$. {\bf Lower graph}: Time-series of variation in species densities (i.e., $\rho_v(t)$) for the experiment shown in the upper graph: horizontal axis is time measured in MCS; vertical axis is variation in density $\rho_v(t)$, as a percentage of $N$.
}
\label{fig:NS07_NA0_L500_T_eg_densities}
\end{figure}

Figure~\ref{fig:NS07_NA0_L500_T_eg_densities} shows an illustrative typical time-series of results for the seven-species system with the unablated dominance network being a tournament graph (i.e., every species $S_i$ dominates three other species, and is dominated by three other species; and all $S_{j\neq i}$ are either dominated by or dominates $S_i$): the upper plot shows the densities $\rho_i(t)$ of the seven species $S_0$ to $S_6$, and the lower plot shows the variation in density $\rho_v(t)$ calculated from the individual densities.  
As can be seen from both graphs, the opening phase of the experiment involves a rapid transition away from the initial conditions, where each species is equally dense (and hence $\rho_v(0)$$\approx$$0$), to a situation in which at any one time some species' densities are low (but will eventually rise) and other species' densities are high (but will eventually fall), the densities rising and falling in a quasiperiodic fashion over the duration of the experiment.  

\begin{figure}
\begin{center}
\includegraphics[trim=0cm 0.1cm 0cm 0cm, clip=true, scale=0.35]{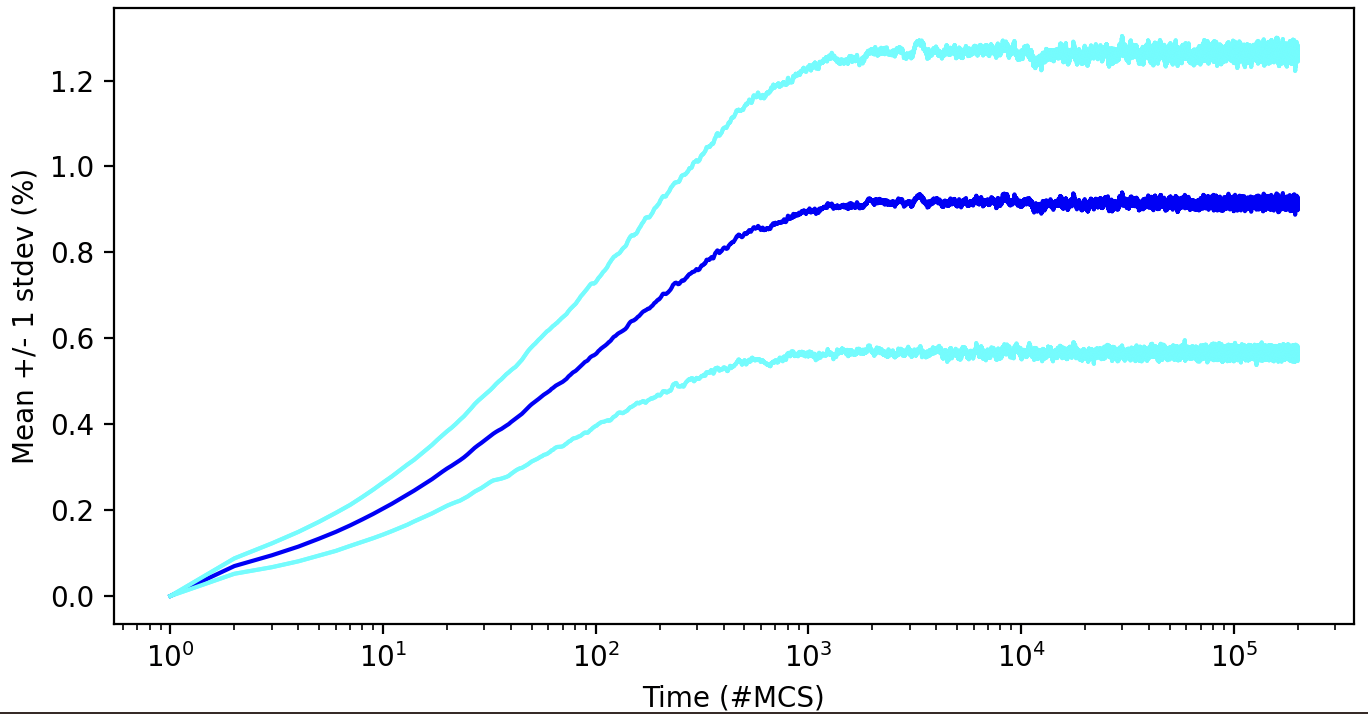}
\end{center}
\caption{Time-series of mean (plus and minus one standard deviation) value of  $\rho_v(t)$ aggregated over $N_S$$=$$7, N_a$$=$$0$ unablated tournament  experiments where $M$ was swept over the range $[10^{-9}, 2$$\times$$10^{-6}]$,  with 100 IID experiments performed at each value of $M$ sampled.  Horizontal axis is time measured in MCS; vertical axis is variation in density $\rho_v(t)$, expressed as a percentage of $N$.
}
\label{fig:NS07_NA0_L500_T_all_densvar}
\end{figure}

\begin{figure}
\begin{center}
\includegraphics[trim=0.6cm 0.8cm 0cm 0.0cm, clip=true, scale=0.325]{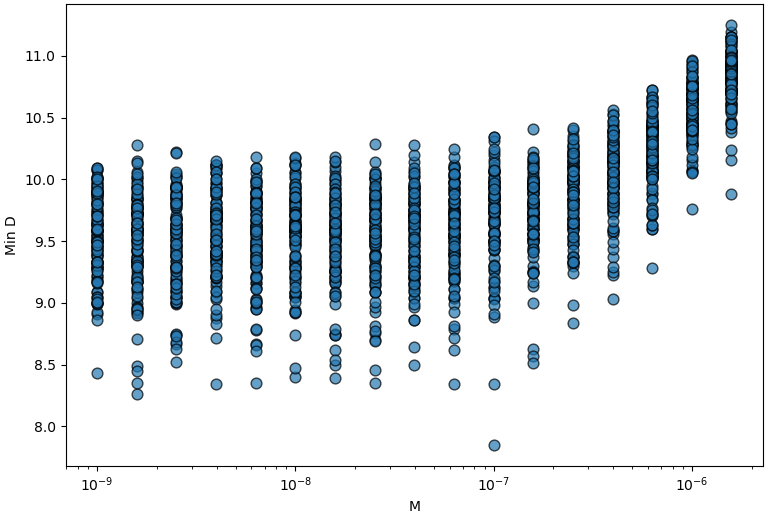}
\end{center}
\caption{Minimum value of  $\rho_i(t) \forall i \forall t$ aggregated over  $N_S$$=$$7, N_a$$=$$0$ unablated tournament  experiments where $M$ was swept over the range $[10^{9}, 2$$\times$$10^{-6}]$,  with 100 IID experiments performed at each value of $M$ sampled.  Horizontal axis is $M$; vertical axis is density, as a percentage of $N$.
}
\label{fig:NS07_NA0_L500_T_all_mindensities}
\end{figure}

To provide a baseline against which the results from ablated-network systems could be judged, a set of experiments were run where the value of $M$ was swept over the range $[10^{-9}, 2$$\times$$10^{-6}]$. At each value of $M$ sampled in this range,  100 IID experiments like the one illustrated in Figure~\ref{fig:NS07_NA0_L500_T_eg_densities} 
were conducted. 
The outcome of every one of these experiments was essentially the same as the one illustrated in Figure~\ref{fig:NS07_NA0_L500_T_eg_densities}: 
the number of species at the end was always seven (i.e., $n_s(10^5)$$=$$7$).  Figure~\ref{fig:NS07_NA0_L500_T_all_densvar} shows the mean and standard deviation of $\rho_v(t)$ aggregated over this entire set of experiments: as can be seen, after the initial transient of roughly 1000MCS, mean $\rho_v(t)$ stabilises at roughly 0.9\% for the remainder of the experiment. Figure~\ref{fig:NS07_NA0_L500_T_all_mindensities} shows the minimum density recorded across all species over all time in each experiment, i.e.\ $\min(\rho_i(t)) \forall$$i \forall$$t$ for the whole set of experiments summarised in Figure~\ref{fig:NS07_NA0_L500_T_all_densvar}:  there is a clear nonlinear relationship between $M$ and the minimum density recorded, which never goes below 7.5\%.

\subsection{Unablated Non-Tournament Circulant Network}
\label{sec:unablated_NTC}

\begin{figure}
\begin{center}
\includegraphics[trim=0cm 0cm 0cm 0cm, clip=true, scale=0.35]{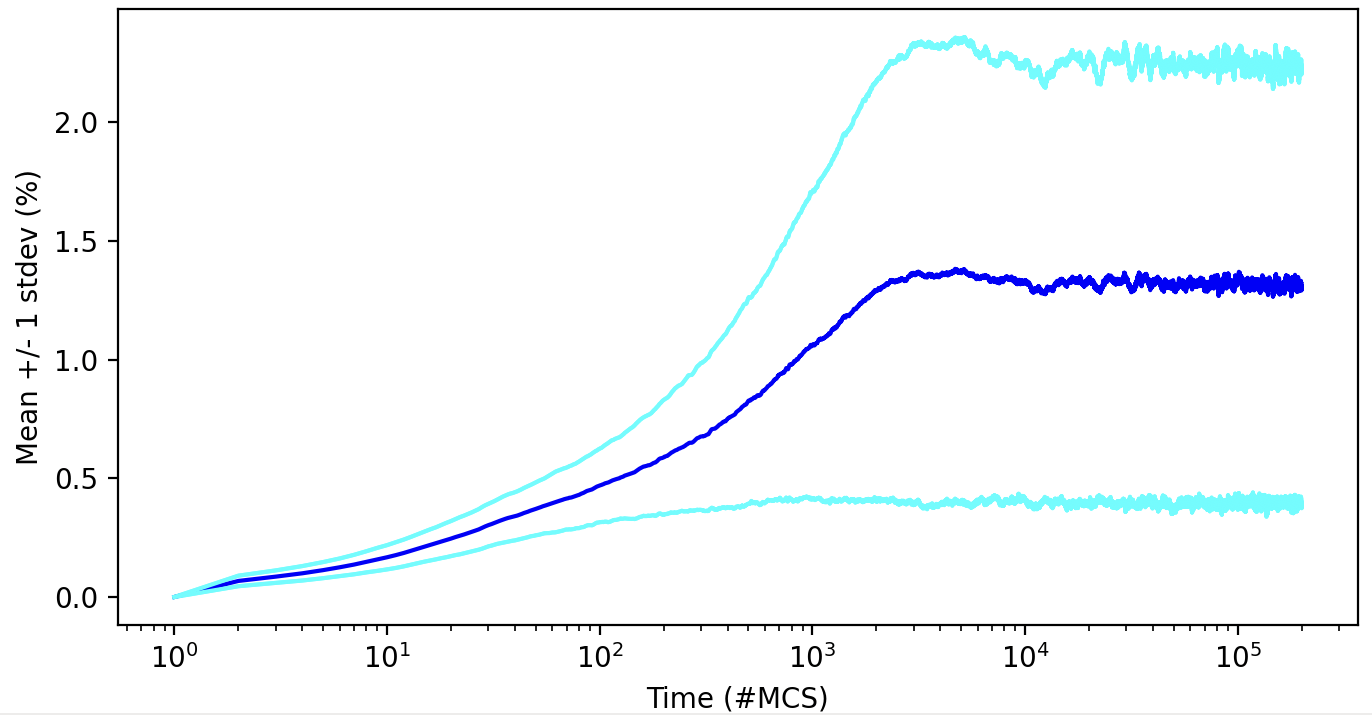}
\end{center}
\caption{Mean ($\pm$$1$ standard deviation) value of  $\rho_v(t)$ aggregated over $N_S$$=$$7, N_a$$=$$0$ unablated NTC experiments where $M$ was swept over the range $[10^{-9}, 2$$\times$$10^{-6}]$,  with 100 IID experiments performed at each value of $M$ sampled.  Format as for Figure~\ref{fig:NS07_NA0_L500_T_all_densvar}.
}
\label{fig:NS07_NA0_L500_C_all_densvar}
\end{figure}

\begin{figure}
\begin{center}
\includegraphics[trim=0cm 0cm 0cm 0cm, clip=true, scale=0.3]{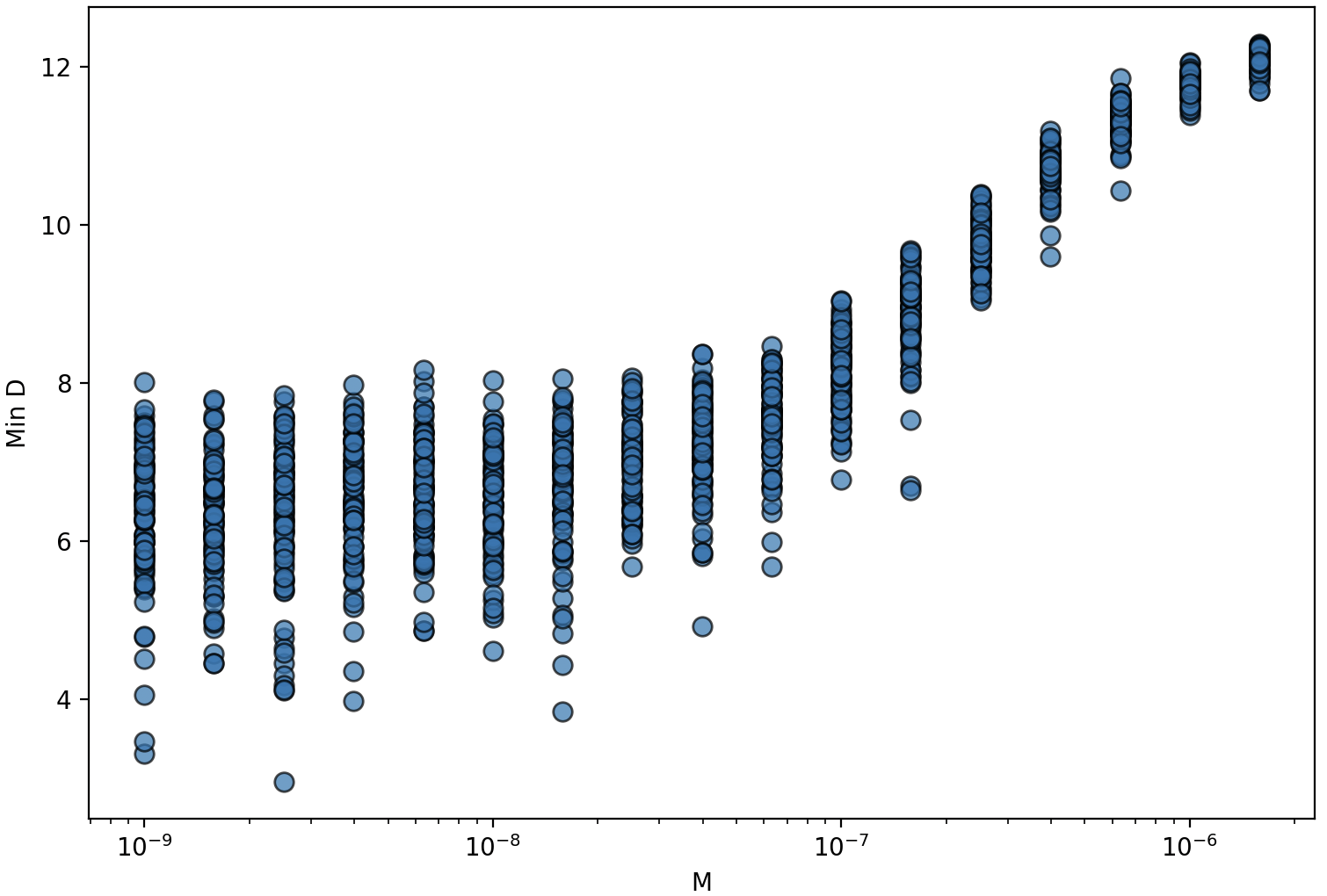}
\end{center}
\caption{Minimum value of  $\rho_i(t) \forall i \forall t$, aggregated over $N_S$$=$$7, N_a$$=$$0$ unablated NTC experiments where $M$ was swept over the range $[10^{-9}, 2\times10^{-6}]$,  with 100 IID experiments performed at each value of $M$ sampled.  Format as for Figure~\ref{fig:NS07_NA0_L500_T_all_mindensities}.
}
\label{fig:NS07_NA0_L500_C_all_mindensities}
\end{figure}

Results from simulations of the ECSG with unablated ($N_a$$=$$0$) non-tournament circulant (NTC) dominance network are qualitatively the same as those from the ECSG with tournament dominance network that were shown in the previous section, but there are some qualitative differences. 
A time-series of the mean $\rho_v(t)$ aggregated over the results from all values of $M$ is shown in Figure~\ref{fig:NS07_NA0_L500_C_all_densvar}: as can be seen, in the NTC system the mean variation in densities is higher, i.e.\ roughly 1.3\% at steady-state, than that of the tournament system, and the standard deviation is wider. Similarly, a scatter-plot of minimum $\rho_i(t)\forall i \forall t$ values recorded in each simulation is shown in Figure~\ref{fig:NS07_NA0_L500_C_all_mindensities}, and the NTC minima are roughly half those of the tournament system at low $M$, while at higher values of $M$ the NTC minima are roughly the same range but more tightly compressed, than those of the tournament-based system illustrated in Figure~\ref{fig:NS07_NA0_L500_T_all_mindensities}.

\subsection{Ablated Tournament Network}
\label{sec:ablated_T}

\begin{figure}
\begin{center}
\includegraphics[trim=0cm 0cm 0cm 0cm, clip=true, scale=0.4]{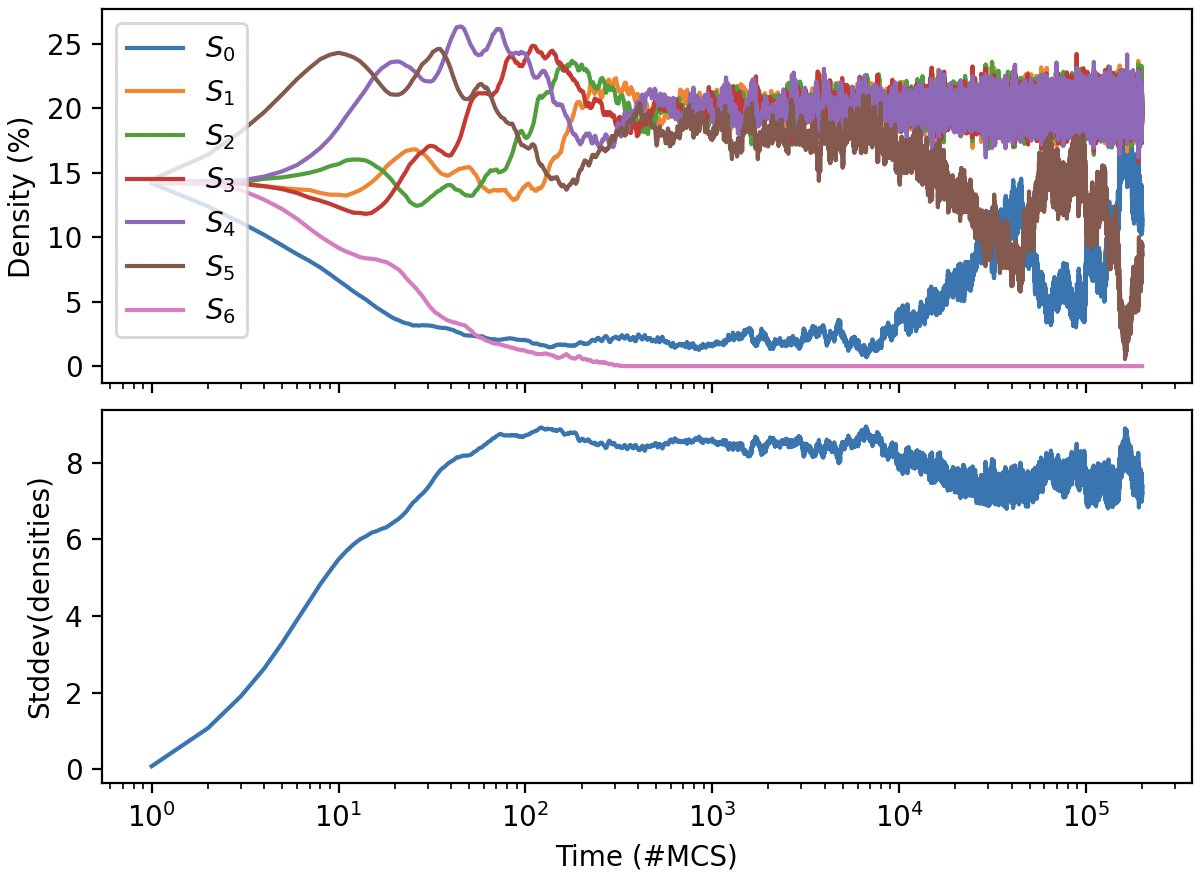}
\end{center}
\caption{Time-series of species densities, and variation in densities, in a single experiment where  one extinction occurred, and hence $n_s(200\text{k})$$=$$6$, in the $N_S$$=$$7, N_a$$=$$1$ ablated tournament system: $L$$=$$500$, $\mu$$=$$\sigma$$=1.0$, $M$$=$$10^{-7}$. Format as for Figure~\ref{fig:NS07_NA0_L500_T_eg_densities}.
}
\label{fig:NS07_NA1_L500_T123_1extinct}
\end{figure}

\begin{figure}
\begin{center}
\includegraphics[trim=0cm 0cm 0cm 0cm, clip=true, scale=0.4]{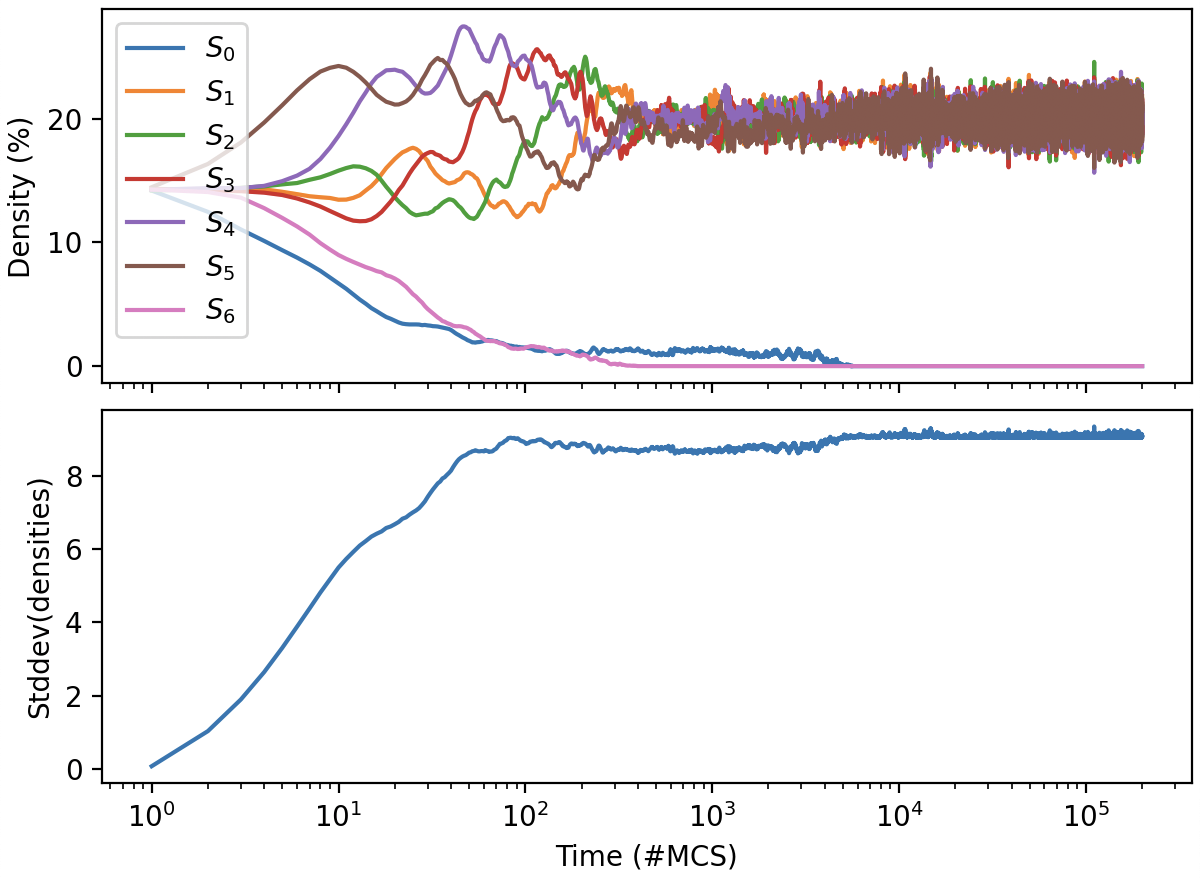}
\end{center}
\caption{Species densities, and variation in densities, in a single experiment where two extinctions extinctions occurred, and hence $n_s(200\text{k})$$=$$5$, in the $N_S$$=$$7, N_a$$=$$1$ ablated tournament system: $L$$=$$500$, $\mu$$=$$\sigma$$=1.0$, $M$$=$$10^{-7}$. Format as for Figure~\ref{fig:NS07_NA0_L500_T_eg_densities}.
}
\label{fig:NS07_NA1_L500_T123_2extinct}
\end{figure}

Illustrative results from a single run of the  2-regular $N_S$$=$$7, N_a$$=$$1$ tournament experiment, where there is one extinction over the course of the experiment and hence $n_s(200\text{k})$$=$$6$ is shown in Figure~\ref{fig:NS07_NA1_L500_T123_1extinct}, and Figure~\ref{fig:NS07_NA1_L500_T123_2extinct} shows results from a different run where two extinctions occurred and hence $n_s(200\text{k})$$=$$5$.

\begin{figure}
\begin{center}
\includegraphics[trim=0cm 0cm 0cm 0.9cm, clip=true, scale=0.4]{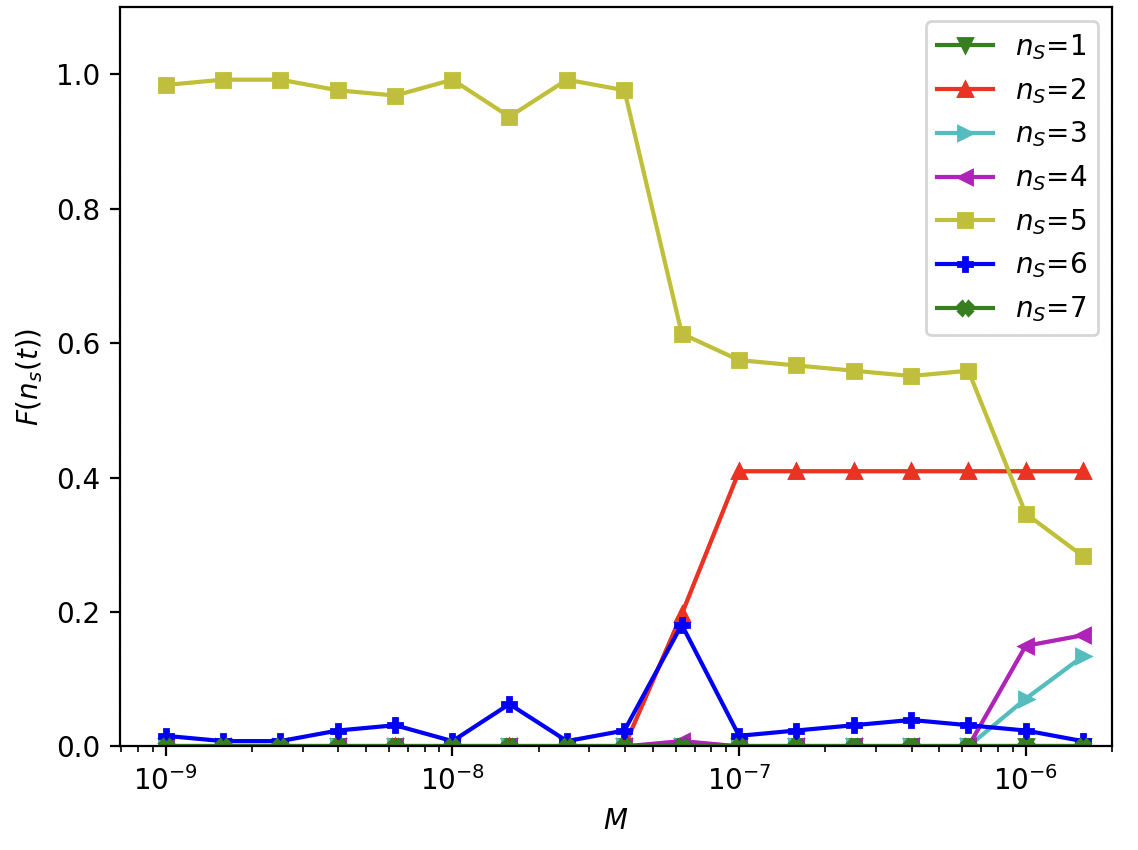}
\end{center}
\caption{Frequency of final species-count at 200kMCS (i.e., $n_s(200\text{k})$) for the $N_s$$=$$7, N_a$$=$$1$ ablated tournament system. 125 IID repetitions were run for each value of $M$. Horizontal axis is $M$; vertical axis is frequency of occurrence of $n_s(200\text{k})$.
}
\label{fig:NS07_NA1_L500_T_FvM}
\end{figure}

\begin{figure}
\begin{center}
\includegraphics[trim=0cm 0.5cm 0cm 1.1cm, clip=true, scale=0.35]{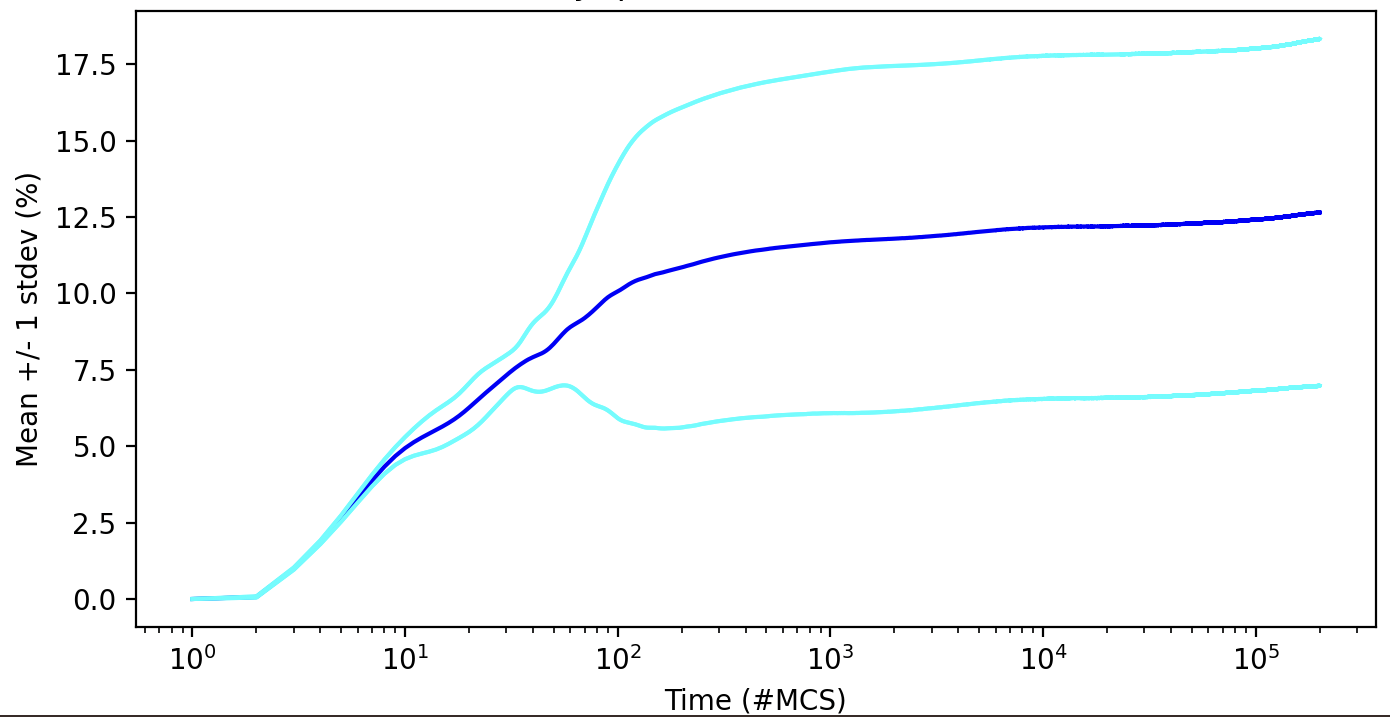}
\end{center}
\caption{Mean ($\pm$$1$ standard deviation) value of  $\rho_v(t)$ aggregated over the 2125 $N_S$$=$$7, N_a$$=$$1$ ablated tournament experiments whose outcomes were summarised in Figure~\ref{fig:NS07_NA1_L500_T_FvM}.
Format as for Figure~\ref{fig:NS07_NA0_L500_T_all_densvar}.
}
\label{fig:NS07_NA1_L500_T_all_densvar}
\end{figure}

Results from the $M$-sweep on 2-regular $N_S$$=$$7, N_a$$=$$1$ tournament experiments are shown in Figure~\ref{fig:NS07_NA1_L500_T_FvM}: as can be seen, for low values of $M$, the outcome $n_s(200\text{k})$$=$$5$ occurs almost 100\% of the time, but as $M$ increases from $4$$\times$$10^{-8}$ to $10^{-7}$ there is a sharp drop in the frequency of 5-species outcomes, falling to 60\% or less, which is matched by a sharp rise in the number of 2-species outcomes (i.e., where  $n_s(200\text{k})$$=$$2$) rising rapidly from the initial 0\% frequency for low $M$ to 40\% for $M$$\geq$$10^{-7}$.  And then, as $M$ is increased beyond $6$$\times$$10^{-6}$, the frequency of 5-species outcomes drops sharply again, falling to roughly 30\%, while the number of 3-species and 4-species outcomes rises. This set of results qualitatively replicates the central finding of Zhong et al.: as $M$ increases, the species biodiversity undergoes a sharp, sudden decline once $M$ is greater than some threshold value. Figure~\ref{fig:NS07_NA1_L500_T_all_densvar} shows the aggregate variation in density for the set of experiments shown in Figure~\ref{fig:NS07_NA1_L500_T_FvM}. Because in every experiment there was at least one extinction, the scatter-plot of minimum $\rho_i(t)$ is a flat-line with all data points on zero, and hence is not plotted here.   

\subsection{Ablated Non-Tournament Circulant Network}
\label{sec:ablated_NTC}

\begin{figure}
\begin{center}
\includegraphics[trim=0cm 0cm 0cm 0cm, clip=true, scale=0.4]{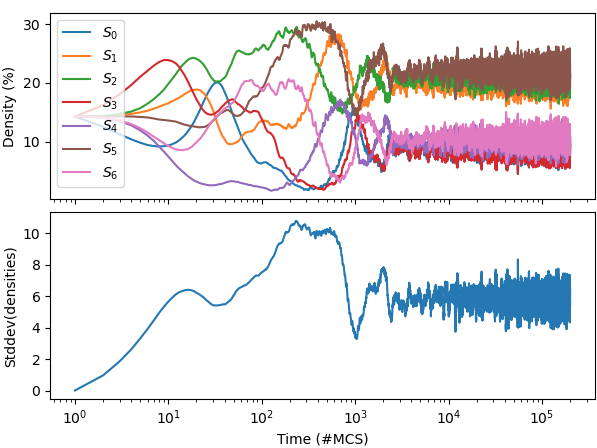}
\end{center}
\caption{Species densities, and variation in densities,  from a single run of the $N_s$$=$$7, N_a$$=$$1$ ablated NTC system where no extinctions occur; $\mu$$=$$\sigma$$=$$1.0$, $M$$=$$10^{-7}$. Format as for Figure~\ref{fig:NS07_NA0_L500_T_eg_densities}.
}
\label{fig:NS07_NA1_L500_C13_0extinct}
\end{figure}

\begin{figure}
\begin{center}
\includegraphics[trim=0cm 0cm 0cm 0cm, clip=true, scale=0.4]{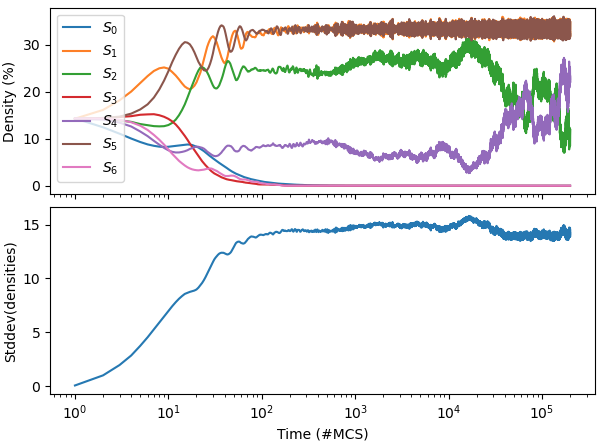}
\end{center}
\caption{Species densities, and variation in densities,  from a single run of the $N_s$$=$$7, N_a$$=$$1$ ablated NTC system where one extinction occurs; $\mu$$=$$\sigma$$=$$1.0$, $M$$=$$10^{-7}$. Format as for Figure~\ref{fig:NS07_NA0_L500_T_eg_densities}.
}
\label{fig:NS07_NA1_L500_C13_3extinct}
\end{figure}

\begin{figure}
\begin{center}
\includegraphics[trim=0cm 0cm 0cm 0cm, clip=true, scale=0.35]{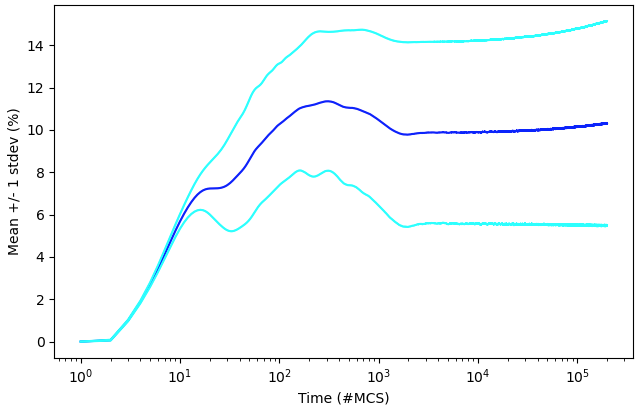}
\end{center}
\caption{Mean ($\pm$$1$ standard deviation) value of  $\rho_v(t)$ aggregated over 1700 $N_S$$=$$7, N_a$$=$$1$ ablated NTC experiments where $M$ was swept over the range $[10^{-9}, 2$$\times$$10^{-6}]$,  with 100 IID experiments performed at each value of $M$ sampled.  Format as for Figure~\ref{fig:NS07_NA0_L500_T_all_densvar}.
}
\label{fig:NS07_NA1_L500_C_all_densvar}
\end{figure}

\begin{figure}
\begin{center}
\includegraphics[trim=0cm 0cm 0cm 0cm, clip=true, scale=0.30]{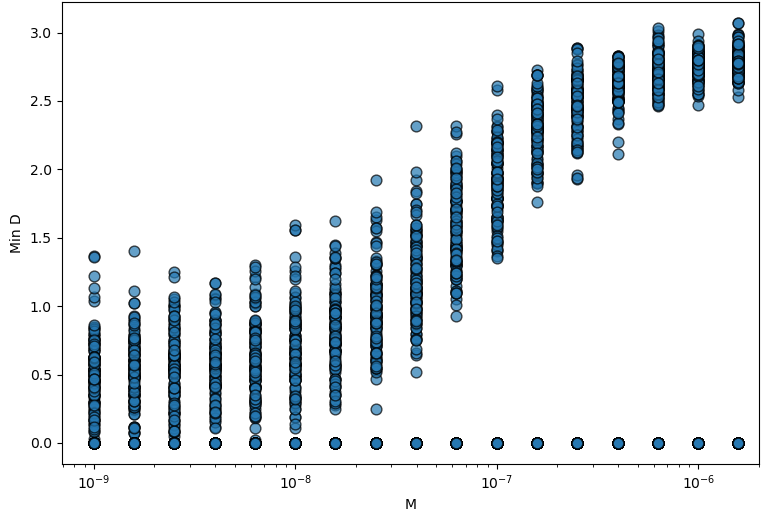}
\end{center}
\caption{Minimum value of  $\rho_i(t), \forall i \forall t$, aggregated over 1700  $N_S$$=$$7, N_a$$=1$ ablated NTC experiments for $M \in [10^{-9}, 2$$\times$$10^{-6}]$,  with 100 IID repetitions at each value of $M$ sampled.  Format as for Figure~\ref{fig:NS07_NA0_L500_T_all_mindensities}.
}
\label{fig:NS07_NA1_L500_C_all_mindensities}
\end{figure}

Typical results from the system with a single random ablation ($N_a$$=$$1$) to the NTC dominance network resulting in no extinctions are shown in Figure~\ref{fig:NS07_NA1_L500_C13_0extinct};
while results from a typical NTC $N_a$$=$$1$ experiment in which three extinctions occur are shown in Figure~\ref{fig:NS07_NA1_L500_C13_3extinct}.
The aggregated $\rho_v$ time-series for this set of experiments is shown in Figure~\ref{fig:NS07_NA1_L500_C_all_densvar}, and the minimum $\rho_i$ data is presented in Figure~\ref{fig:NS07_NA1_L500_C_all_mindensities}.

\begin{figure}
\begin{center}
\includegraphics[trim=0cm 0cm 0cm 0cm, clip=true, scale=0.4]{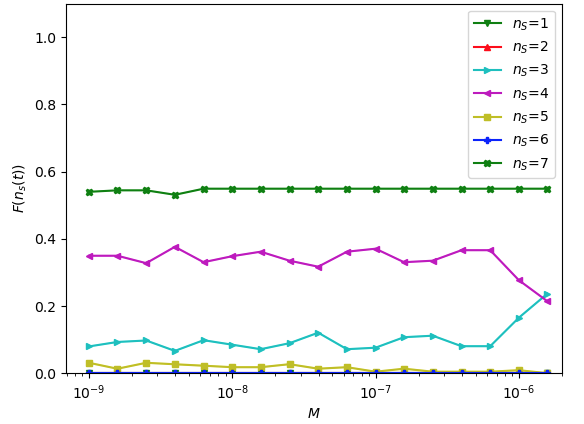}
\end{center}
\caption{Frequency of final species-count at 200kMCS (i.e., $n_s(2$$\times$$10^5)$) for the $N_S$$=$$7, N_a$$=$$1$ ablated NTC system. 200 IID simulations performed for each value of $M$. Format as for Figure~\ref{fig:NS07_NA1_L500_T_FvM}.
}
\label{fig:NS07_NA1_L500_C_FvM}
\end{figure}

Figure~\ref{fig:NS07_NA1_L500_C_FvM} shows the frequency of outcomes as the value of $M$ is swept across its range: these results from the $M$-sweep 2-regular $N_S$$=$$7, N_a$$=$$1$ NTC experiments differed markedly from those of the 3-regular $N_S$$=$$7, N_a$$=$$1$ tournament experiments, in that for the NTC systems roughly 55\% of the IID repetitions showed no extinctions at all, with $n_s(200\text{k})$$=$$N_S$$=$$7$, and the frequencies of $n_s(200\text{k})$$=$$3$ and $n_s(200\text{k})$$=$$4$ outcomes are also essentially constant with respect to variation in $M$ over most of its range, albeit with some changes occurring once $M$$\geq$$6$$\times$$10^{-7}$.
 
\section{Discussion and Conclusion}
\label{sec:discussion}

My results presented in Figure~\ref{fig:NS07_NA1_L500_T_FvM} show that the ESCG simulator I have written for this study does qualitatively reproduce the central result of  \cite{zhong_etal_2022_ablatedRPSLS}: in the specific case of a seven-species tournament RPSLS-like system, deleting a single directed edge ($N_a$$=$$1$) can change the system's overall population dynamics such that, as mobility $M$ is increased beyond some threshold value (here, $4$$\times$$10^{-8}$) there is a sharp fall-off in the frequency of experiment outcomes where the system settles to five-species coexistence, counterbalanced by a sharp rise in the frequency of experiments where the system settles to two-species coexistence: this is the collapse in biodiversity witnessed in the five-species RPSLS system reported by   \cite{zhong_etal_2022_ablatedRPSLS}. In one sense, the results of Figure~\ref{fig:NS07_NA1_L500_T_FvM} support those of \cite{zhong_etal_2022_ablatedRPSLS}, because the biodiversity collapse has now been demonstrated both in the five-species tournament $D(5,\{1,3\})$ system and in the seven-species $D(7,\{1,3,5\})$ system. 
 
However, my results presented in Figure~\ref{fig:NS07_NA1_L500_C_FvM} cast significant doubt on the generality of Zhong et al.'s result: 
surely, for \cite{zhong_etal_2022_ablatedRPSLS} to be of genuine scientific interest, to actually be of relevance to real ecosystems biodiversity, such biodiversity collapses should be seen in any, or at least many, plausible ``interaction structures'' (i.e., dominance networks), but Figure~\ref{fig:NS07_NA1_L500_C_FvM} shows that a manifestly closely related interaction structure, the non-tournament circulant network $D(7,\{1,3\})$ does not undergo the sudden, phase-transition-like, biodiversity collapse for any plausible value of $M$. Thus, it seems reasonable to conjecture that the {\em prima facie} interesting results of \cite{zhong_etal_2022_ablatedRPSLS} are in fact nothing more than a quirk arising from their choice to study only the one RPSLS interaction structure, the tournament circulant $D(5,\{1,3\})$: I have shown here that if they had searched for the same phenomenon in only ever so slightly different interaction structures, they would not have found it.

A counterargument to what I am concluding here could potentially be based on claiming that tournament interaction structures, where every single species is either predator or prey to every other species in the entire ecosystem, are somehow more realistic models of actual biological ecosystems than are circulant interactions structures, where each species is predator/prey to only a (potentially small) subset of the entire array of species within the ecosystem. But such a counterargument could be quickly defeated by anyone armed with junior-school understanding of real biological food webs: the shrimp eat the plankton; the small fish species feed on the shrimp; the medium-size fish species eat the small fish; the big predator fish species eat the medium-size fish; and the bird of prey dives to dine on the big predator fish -- the point here being that eagles don't eat plankton, or shrimp. My view is that the non-tournament circulant (NTC) interaction structure ESCGs  explored here, apparently for the first time, are potentially more realistic than the ECSGs using tournament networks such as RPS and RPSLS that so many researchers have focused exclusively on for so long.

To some extent, it seems that the near-exclusive focus in the literature on tournament-structured RPS and RPSLS has been motivated by a desire to do what \cite{kuhn_1962} referred to as {\em normal science}: experimenting and observing and theorizing within the intellectual confines of a settled, agreed explanatory framework. As Isaac Newton famously wrote to Robert Hooke, we see further when we stand on the shoulders of giants:  if some researcher ${\cal R}$ publishes a ground-breaking and influential paper about a model system ${\cal M}$ then it is natural, it is sound scientific practice, for some other researcher ${\cal S}$ to replicate and extend ${\cal R}$'s study of ${\cal M}$, and then for another researcher ${\cal T} $ to replicate and extend ${\cal S}$'s work on ${\cal M}$, which may then prompt researcher ${\cal U}$ to replicate and extend  ${\cal T}$'s study, and so on and so forth, and hence ${\cal M}$ can become locked-in, becoming one of the prevailing paradigm's standard models that everyone in the field is familiar with and accepting of, and which is studied to the exclusion of other equally (or perhaps more) plausible and relevant models that in fact the original researcher ${\cal R}$ could just as easily have chosen to work on in the first place, but just happened not to.

There are many directions in which the work reported here could be taken further, several of which were mentioned in passing in the main text of this paper. The Python source-code used for the simulations reported here is freely available as open-source on GitHub, under the MIT Open-Source License,\footnote{See {\tt https://github.com/davecliff/ESCG\_Python}.} to allow other researchers to readily replicate and extend this work.

\bibliographystyle{./apa}

\bibliography{../../dc_bibliography}

\section*{Appendix: Three Modes of Response}

\begin{figure}[t]
\begin{center}
\includegraphics[trim=0cm 0cm 0cm 0cm, clip=true, scale=0.34]{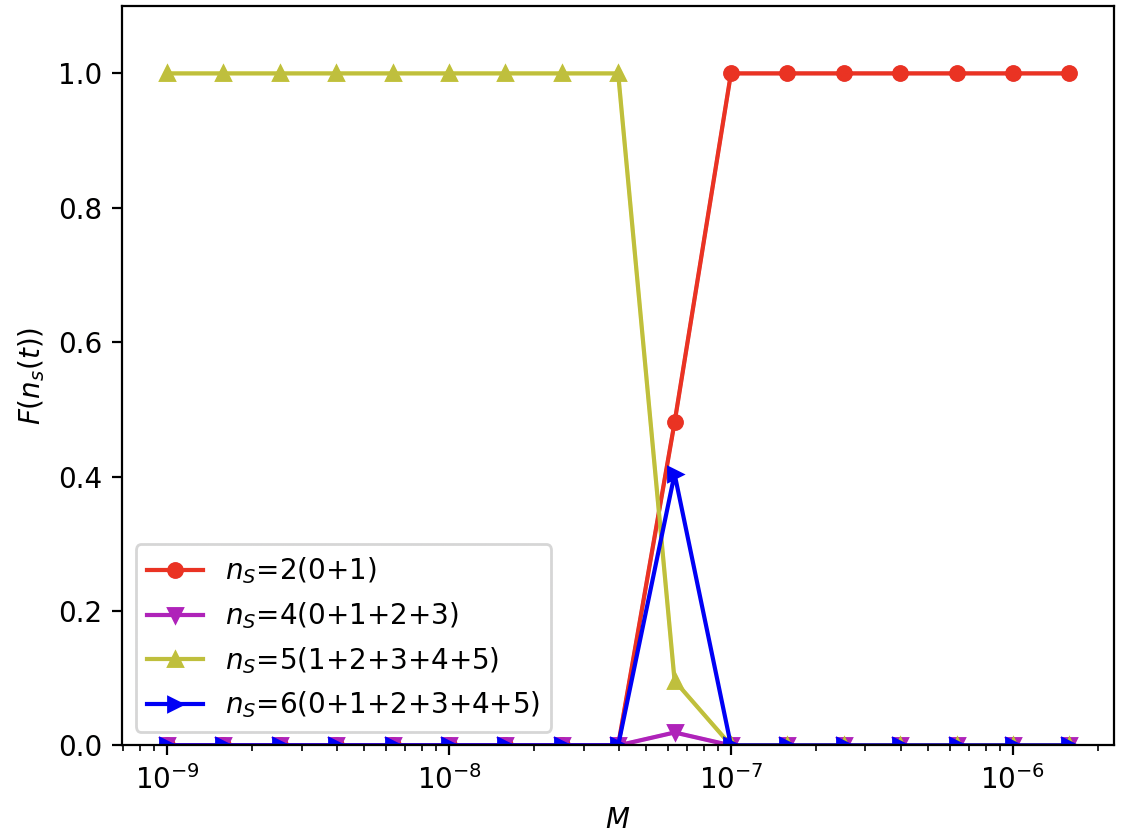}
\end{center}
\caption{Frequency of final species-count at 200kMCS (i.e., $n_s(200\text{k})$) for the $N_s$$=$$7, N_a$$=$$1$ ablated tournament system when the ablated edge is the dominance interaction from $S_0$$\rightarrow$$S_1$. Graph shows results from approx.\ 800 IID repetitions in total, so roughly 50 for each value of $M$. Format as for Figure~\ref{fig:NS07_NA1_L500_T_FvM}, but here the legend also shows, in parenthesis, the indexes of the surviving species. 
}
\label{fig:NS07_NA1_L500_T_FvM_0001010}
\end{figure}

\begin{figure}
\begin{center}
\includegraphics[trim=0cm 0cm 0cm 0cm, clip=true, scale=0.34]{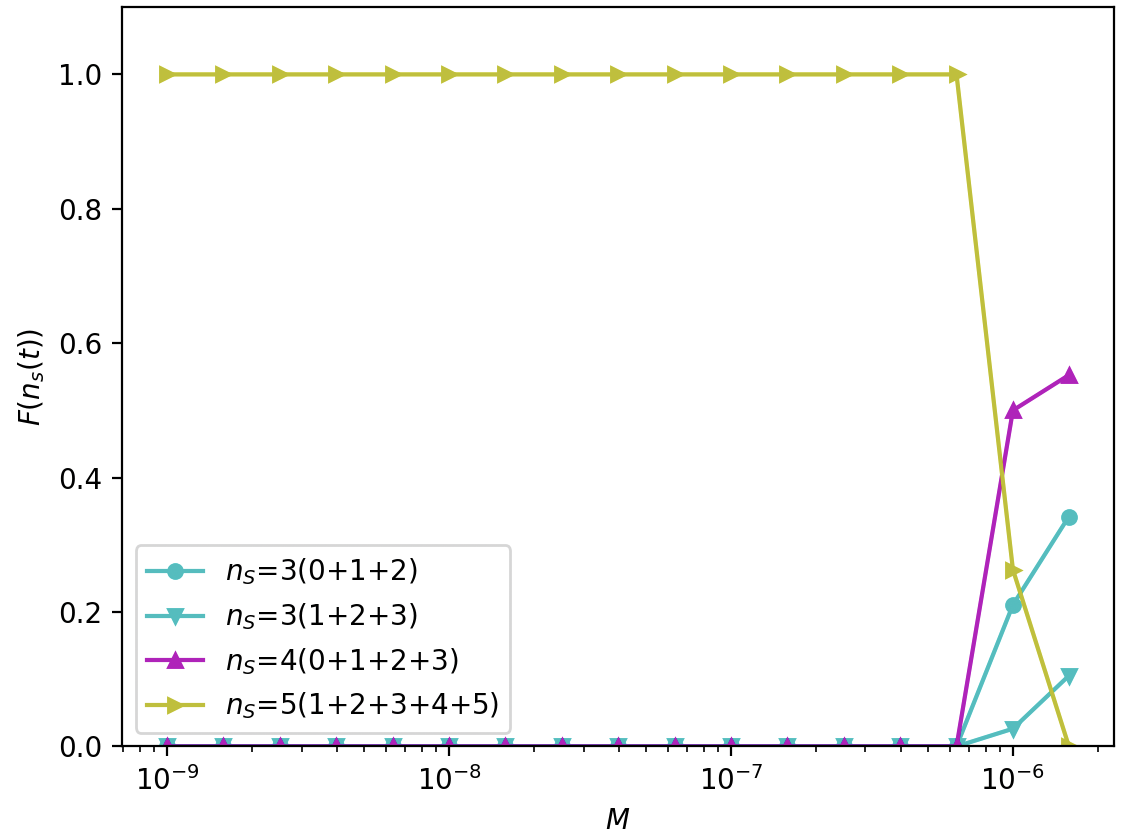}
\end{center}
\caption{Frequency of final species-count at 200kMCS (i.e., $n_s(200\text{k})$) for the $N_s$$=$$7, N_a$$=$$1$ ablated tournament system when the ablated edge is the dominance interaction from $S_0$$\rightarrow$$S_3$. 
Graph shows results from approx.\ 800 IID repetitions in total, so roughly 50 for each value of $M$. 
Format as for Figure~\ref{fig:NS07_NA1_L500_T_FvM_0001010}. 
}
\label{fig:NS07_NA1_L500_T_FvM_0100010}
\end{figure}

\begin{figure}
\begin{center}
\includegraphics[trim=0cm 0cm 0cm 0cm, clip=true, scale=0.34]{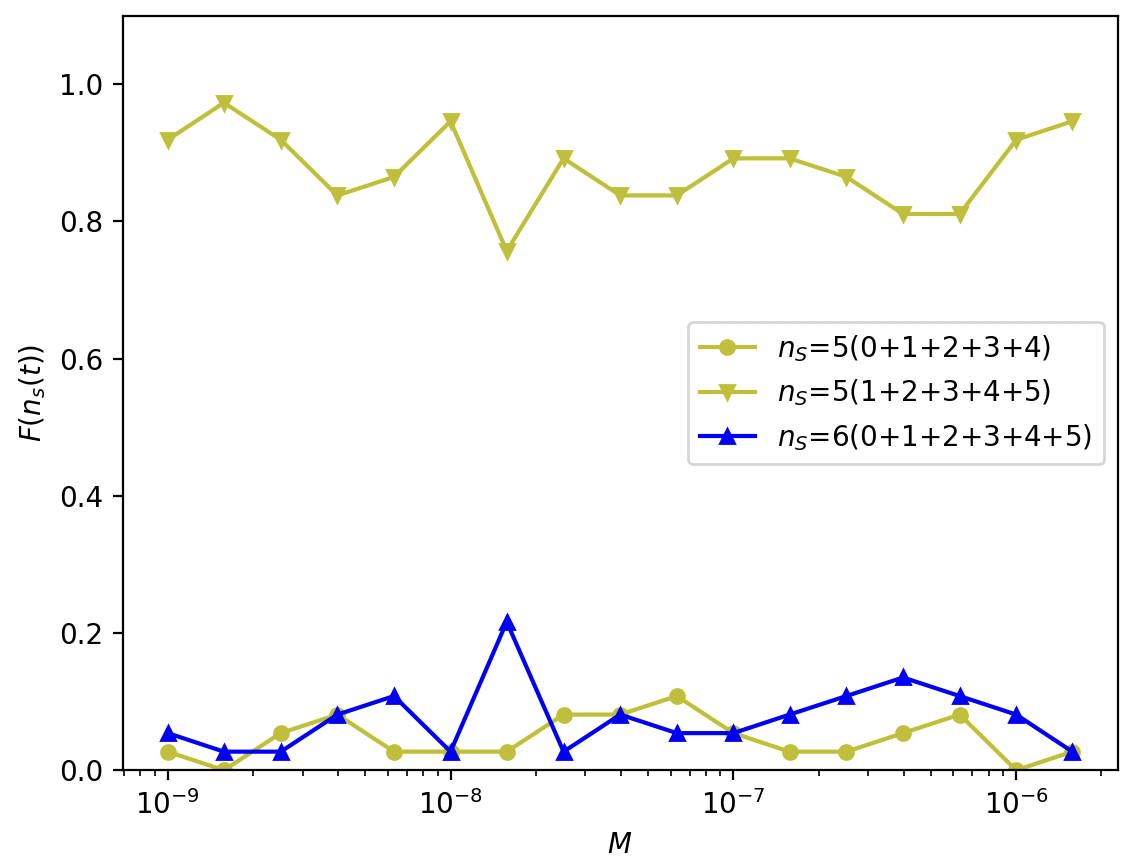}
\end{center}
\caption{Frequency of final species-count at 200kMCS (i.e., $n_s(200\text{k})$) for the $N_s$$=$$7, N_a$$=$$1$ ablated tournament system when the ablated edge is the dominance interaction from $S_0$$\rightarrow$$S_5$. 
Graph shows results from approx.\ 800 IID repetitions in total, so roughly 50 for each value of $M$. 
Format as for Figure~\ref{fig:NS07_NA1_L500_T_FvM_0001010}.
}
\label{fig:NS07_NA1_L500_T_FvM_0101000}
\end{figure}

Figure~\ref{fig:NS07_NA1_L500_T_FvM} summarised results from approx 2,500 tournament experiments where in each experiment one of the dominance relationships from species $S_0$ to species $S_1$, $S_3$, and $S_5$ was randomly 
 for deletion, and as such it presented the superposition of three different modes of response. For completeness, Figures~\ref{fig:NS07_NA1_L500_T_FvM_0001010}, \ref{fig:NS07_NA1_L500_T_FvM_0100010}, and~\ref{fig:NS07_NA1_L500_T_FvM_0101000} show the frequency of outcome for each of those three modes, separately. 
A corresponding set of three graphs could be plotted for the data  shown in 
Figure~\ref{fig:NS07_NA1_L500_C_FvM}, but the lack of variation in that data makes for three boring graphs, not shown here.

\end{document}